%% file: arxiv.tex
\documentclass[aps,prx,twocolumn,longbibliography,floatfix,10pt,superscriptaddress]{revtex4-1}
\usepackage[utf8]{inputenc}
\usepackage{natbib}
\usepackage{graphicx}
\usepackage{latexsym}
\usepackage{amssymb}
\usepackage{amsmath}
\usepackage{amsfonts}
\usepackage{gensymb}
\usepackage{bm}
\usepackage{hyperref}
\usepackage{braket}
\usepackage{physics}
\usepackage{bbold}
\usepackage[caption=false]{subfig}
\usepackage[dvipsnames]{xcolor}
\usepackage[normalem]{ulem}
\usepackage{siunitx}
\usepackage{mhchem}
\usepackage{multirow}
\usepackage{slashed}
\usepackage{comment}

\newcommand{\PRLsec}[1]{\emph{#1---}}

\newcommand{\supplementarysection}{%
  \setcounter{figure}{0}
  \let\oldthefigure\thefigure
  \renewcommand{\thefigure}{S\oldthefigure}
  \setcounter{section}{0}
  \let\oldthesection\thesection
  \renewcommand{\thesection}{S\oldthesection}
  \setcounter{equation}{0}
  \let\oldtheequation\theequation
  \renewcommand{\theequation}{S\oldtheequation}
  \setcounter{table}{0}
  \let\oldthetable\thetable
  \renewcommand{\thetable}{S\oldthetable}
}

\newcommand{\bi}{\begin{itemize}}
\newcommand{\ei}{\end{itemize}}
\newcommand{\be}{\begin{enumerate}}
\newcommand{\ee}{\end{enumerate}}

\newenvironment{dfn}{{\vspace*{1ex} \noindent \bf Definition }}{\vspace*{1ex}}

\newcommand{\br}{\mathbf{r}}
\newcommand{\bk}{\mathbf{k}}

	\newcommand{\beq}{\begin{eqnarray}}
	\newcommand{\eeq}{\end{eqnarray}}

	\renewcommand{\v}{{\bf v}}

\begin{document}

\title{Emergent QED$_3$ at the bosonic Laughlin state to superfluid transition}

\author{Taige Wang}
\affiliation{Department of Physics, University of California, Berkeley, CA 94720, USA \looseness=-2}
\affiliation{Kavli Institute for Theoretical Physics, University of California, Santa Barbara, CA 93106, USA \looseness=-2}
\author{Xue--Yang Song}
\affiliation{Department of Physics, Hong Kong University of Science and Technology, Clear Water Bay, Hong Kong 999077, China \looseness=-2}
\affiliation{Department of Physics, Massachusetts Institute of Technology, Cambridge, Massachusetts 02139, USA \looseness=-2}
\author{Michael P. Zaletel}
\affiliation{Department of Physics, University of California, Berkeley, CA 94720, USA \looseness=-2}
\affiliation{Material Science Division, Lawrence Berkeley National Laboratory, Berkeley, CA 94720, USA \looseness=-2}
\author{T. Senthil}
\affiliation{Department of Physics, Massachusetts Institute of Technology, Cambridge, Massachusetts 02139, USA \looseness=-2}


\begin{abstract}

Quantum phase transitions between  topologically ordered and symmetry-broken phases lie beyond Landau theory.  
A prime example is the conjectured continuous transition from the bosonic \(\nu = 1/2\) Laughlin state to a superfluid, proposed to be governed by a QED\(_3\)–Chern–Simons (CS) critical point whose stability remains uncertain.  
We study half-filled bosons in the lowest Landau level subject to a lattice potential. Infinite-cylinder DMRG reveals a single continuous Laughlin–to–superfluid transition.  
Adiabatic flux insertion collapses the many-body gap and exposes massless Dirac quasiparticles, while momentum-resolved correlation lengths show that three lattice-related density modes share the same critical exponent, evidencing an emergent \(SO(3)\) symmetry.  
The joint appearance of Dirac dispersion and symmetry enlargement provides microscopic support for a stable QED\(_3\)–CS fixed point.  
Our numerical strategy offers a blueprint for exploring Landau-forbidden transitions in fractional Chern insulators and composite Fermi liquids realised in moiré and cold-atom systems.

\end{abstract}

\maketitle

The best studied examples of continuous quantum phase transitions involve  the onset of broken symmetry from a trivial gapped symmetry preserving phase~\cite{sachdev1999quantum}. These  are described by the Landau-Ginzburg-Wilson (LGW) paradigm which dictates that the universal critical properties are captured by a continuum quantum field theory for the order parameter field. This paradigm generally fails~\cite{senthil2024deconfined} for other kinds of quantum phase transitions. For instance, either phase may itself host non-Landau order that eludes any conventional order parameter description~\cite{wen2004quantum}. Moreover, even when both phases are individually Landau-allowed, the transition between them can still be continuous and beyond Landau due to the presence of emergent gauge fields~\cite{DQCP1,DQCP2,DQCP3}.

Here we focus on continuous phase transitions that connect a topologically ordered phase to a symmetry breaking phase. A simple pathway is \emph{anyon condensation}~\cite{burnell2018anyon}: when a bosonic anyon that braids non-trivially with all other anyons and carries a microscopic symmetry charge condenses, its condensation simultaneously confines the remaining anyons and breaks the symmetry~\cite{XY_str1,jalabert1991spontaneous,senthil2000z}. A recent example is the transition from the bilayer $\nu = 1/3 + 2/3$ fractional quantum Hall state to an exciton superfluid. It is driven by condensation of the fractional exciton $(e/3,-e/3)$ and therefore belongs to the XY$^{*}$ universality class~\cite{XY*}.

The phase transition we investigate lies beyond the standard anyon condensation mechanism, yet it may still be continuous~\cite{kivelson1992global,chen1993mott,wen1993transitions,MaissamLaughlin,JYL,song_deconfined2023,song_2024_transition}. Specifically, we study the transition between the bosonic $\nu = 1/2$ FQH state to a superfluid in $(2+1)$-dimension. The superfluid breaks the microscopic $U(1)$ symmetry, whereas the $\nu = 1/2$ Laughlin state exhibits topological order without symmetry breaking and hosts a single gapped semion. Because no bosonic anyon is available to condense, the usual anyon-condensation route cannot connect the two phases.

Nonetheless, a field theory for a continuous transition between them has been proposed~\cite{MaissamLaughlin,song_deconfined2023,song_2024_transition}. One formulation employs $N_f = 2$ massless Dirac fermions coupled to a $U(1)$ gauge field with a Chern–Simons term (QED$_3$-CS). Early work suggested that the superfluid–Laughlin transition required point-group symmetry (e.g., inversion)~\cite{MaissamLaughlin}, but later studies showed that, lattice translation symmetry alone suffices at half filling~\cite{song_deconfined2023,song_2024_transition}. The same QED$_3$-CS theory also underpins other experimentally relevant transitions, such as between a Fermi liquid and a composite Fermi liquid~\cite{MaissamCFL,song_2024_transition}. Whether the $N_f = 2$ QED-CS truly governs the Laughlin–superfluid critical point and, more broadly, whether it flows to a conformal fixed point remains unresolved, making a numerical study of this transition crucial~\footnote{A $1/N_f$ expansion predicts conformality, but its validity at $N_f = 2$ is uncertain.}.

Previous numerical studies, largely on lattice hopping models~\cite{NormLaughlin,NormChern,Frank2017,Zeng2021}, have not settled the issue. The transition has alternately appeared first-order, indirect, or inaccessible.  
In this work, we revisit the problem in a continuum setting: half-filled bosons confined to the lowest Landau level and perturbed by a lattice potential.  
For a weak potential the ground state is the bosonic \(\nu = 1/2\) Laughlin fluid; for strong potential kinetic energy dominates and a superfluid emerges despite the magnetic field.  
Using infinite-cylinder density-matrix renormalisation group (DMRG) we find a single, continuous Laughlin–to–superfluid transition.  

We show that adiabatic insertion of a \(2\pi\) flux quantum through the cylinder collapses the many-body gap linearly and exposes a massless Dirac spectrum, as expected at a QED$_3$ critical point.  
Simultaneously, momentum-resolved correlation lengths reveal that three lattice-related charge-density operators acquire identical critical exponents, signaling an emergent \(\mathsf{SO(3)}\) symmetry predicted by the QED$_3$–CS theory.  
The joint appearance of a vanishing Dirac gap and symmetry enlargement therefore furnishes microscopic evidence for the \emph{stability} of the QED$_3$–CS fixed point.  

Beyond resolving a longstanding question about the Laughlin–superfluid transition, our work introduces flux-insertion spectroscopy and symmetry-resolved scaling as practical diagnostics for Landau-forbidden criticality.  
These tools can be applied to fractional Chern insulators, composite Fermi liquids, and other topological phases now realized in moiré heterostructures and cold-atom platforms.

\begin{figure}[t]
    \centering
    \includegraphics[width=0.95\linewidth]{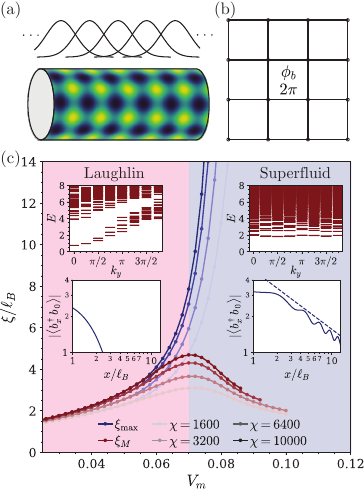}
    \caption{%
    (a)~Setup: bosons in the LLL on an infinite cylinder subject to a square lattice potential.  
    (b)~The unit cell of the physical boson encloses one flux quantum (\(\phi_b=2\pi\)).  
    (c)~Phase diagram at \(L_y=8a\).  The longest neutral correlation length \(\xi_{\max}\) (blue) and the \(M\)-point correlation length \(\xi_M\) (red) are shown for several bond dimensions \(\chi\). Their behaviour identifies a single Laughlin–to–superfluid transition near \(V_m\approx0.07\). Insets give entanglement spectra (top) and real-space correlators (bottom) on either side of the transition (\(V_m = 0.02\) and \(0.1\) respectively).}
    \label{fig:pd}
\end{figure}

\PRLsec{Bosons in the LLL with a periodic potential}  
We consider bosons confined to the lowest Landau level (LLL) on an infinite cylinder, Fig.~\ref{fig:pd}(a).  
The particles feel a square-lattice potential and a contact repulsion  
\begin{equation}
\label{eq:H_def}
H=\frac12 \int_{\br,\br'}\hat\rho_{\br}\delta_{\br-\br'}\hat\rho_{\br'}
  -2\pi V_m\sum_{j=1}^{2}\int_{\br}\hat\rho_{\br}\cos\bigl(\mathbf G_j\cdot\br\bigr),
\end{equation}
where \(\hat\rho_{\br}\) is the \textit{bare} (unprojected) density operator and \(\mathbf G\) are the reciprocal lattice vectors. The lattice spacing \(a=\sqrt{2\pi}\ell_B\) places one flux quantum in every plaquette as shown in Fig.~\ref{fig:pd}(b). For a triangular lattice we obtain the same phase diagram and critical behavior as detailed in the Supplementary Material~\cite{SM}.

At \(V_m=0\) Hamiltonian~\eqref{eq:H_def} reduces to the parent Hamiltonian of the bosonic \(\nu=1/2\) Laughlin state, which is an exact zero-energy ground state~\cite{ParentHamiltonian}.  
In the opposite limit \(V_m\to\infty\), the potential dominates and yields the single-particle dispersion  
\begin{equation}
\label{eq:dis}
\varepsilon_{\mathbf k}=-2\pi e^{-\pi/2}V_m
  \sum_{j=1}^{2}\cos\bigl(\mathbf k\cdot\mathbf a_j\bigr),
\end{equation}
with lattice primitives \(\mathbf a\)'s.  
The dispersion minimum sits at \(\Gamma\). Therefore, bosons condense into a zero-momentum superfluid.

To access intermediate \(V_m\) we solve Eq.~\eqref{eq:H_def} with infinite-cylinder DMRG, conserving both the global \(U(1)\) charge and the longitudinal translation symmetry~\cite{ExactMPS,TopoChara}.  
For circumferences \(L_y=6a\)–\(12a\) we find a single continuous transition. Fig.~\ref{fig:pd}(c) shows the result for \(L_y=8a\), where \(V_m^{c}\simeq 0.07\)~\footnote{For very narrow cylinders the Laughlin state evolves into a charge-density wave~\cite{lee_abelian}. At the larger \(L_y\) used here the entanglement spectrum confirms genuine Laughlin order.}.

The charge-neutral correlation length offers a first diagnostic for the phase diagram.  
It is finite in the gapped Laughlin phase and diverges in the gapless superfluid due to the Goldstone mode.  
As Fig.~\ref{fig:pd}(c) shows, the maximal neutral length \(\xi_{\max}\) rises steeply as \(V_m\) approaches \(V_m^{c}\) and remains large for \(V_m>V_m^{c}\).  
A sharper indicator is \(\xi_{M}\), the correlation length associated with operators at the \(M\) point (see later sections for definition and numerical extraction).  
\(\xi_{M}\) peaks \emph{only} at criticality, providing a precise numerical estimate of \(V_m^{c}\).

Further confirmation of the phase diagram comes from the entanglement spectrum and the single-boson Green’s function  
\(G(r)\equiv\langle b^{\dagger}_{\mathbf r}b_{0}\rangle\).  
For \(V_m<V_m^{c}\) the entanglement spectrum displays the \((1,1,2,3,\dots)\) chiral-boson counting expected for a \(\nu=1/2\) Laughlin edge, consistent with the Li–Haldane conjecture~\cite{Li-Haldane} (upper-left inset of Fig.~\ref{fig:pd}(c)).  

In the thermodynamic limit a 2D superfluid possesses off-diagonal long-range order: \(G(r)\) approaches a non-zero constant at large \(r\) even when the ground state formally preserves \(U(1)\) symmetry.  
Our iDMRG simulations, however, are carried out on a quasi-1D cylinder. The Mermin–Wagner theorem forbids spontaneous breaking of the continuous \(U(1)\) symmetry, and quasi-1D superfluid exhibits \emph{algebraic long-range order},  
\begin{equation}
G(r)\sim r^{-\eta}
\end{equation}
which is clearly visible for \(V_m > V_m^c\) (lower-right inset of Fig.~\ref{fig:pd}(c)).  By contrast, the Laughlin phase exhibits exponential decay set by its bulk gap.  

The divergent correlation lengths at \(V_m^{c}\), Laughlin-edge entanglement spectra for \(V_m<V_m^{c}\), and algebraic long-range order for \(V_m>V_m^{c}\) demonstrates that the Hamiltonian in Eq.~\eqref{eq:H_def} undergoes a single, continuous transition from the bosonic Laughlin state to a superfluid without any intermediate phase.

\begin{figure}[t]
    \centering
    \includegraphics[width=\linewidth]{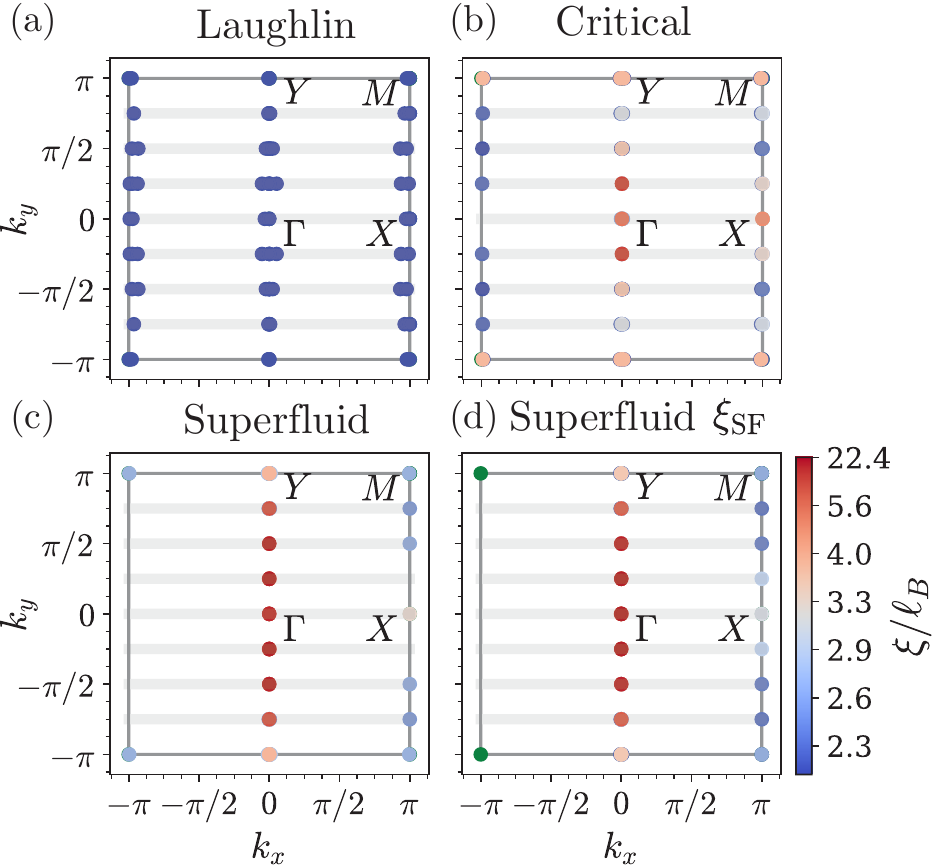}
    \caption{Momentum–resolved correlation lengths \(\xi(\mathbf k)\) (color scale) on a cylinder of circumference \(L_y=8a\) at \(\chi = 10000\). Panels (a)–(c) show the charge-neutral sector at \(V_m=0.022\) (Laughlin), \(0.084\) (critical), and \(0.10\) (superfluid). Panel (d) shows the charge-$1$ sector at \(V_m=0.10\).  At criticality the correlation lengths at \(X\), \(Y\), and \(M\) become almost equal, indicating an emergent \(\mathsf{SO(3)}\) symmetry.}
    \label{fig:xi_k}
\end{figure}

\PRLsec{Phase transition and critical theory}  
To unravel how correlations evolve across the transition we decompose any operator as \(O(\br)=\sum_{\bk\in\mathrm{BZ}} e^{i\bk\cdot\br} O_{\bk}(\br)\) and track the correlation length of each slowly varying component \(O_{\bk}(\br)\). The numerical procedure to extract these lengths from the transfer matrix is detailed in the Supplementary Material~\cite{SM}. Fig.~\ref{fig:xi_k} presents the resulting momentum-resolved correlation lengths. In the charge-neutral sector, all correlation lengths are short deep inside the Laughlin phase, whereas in the superfluid the dominant correlation appears at the Brillouin-zone center \(\Gamma\), as expected from the Goldstone mode.  In the charged sector the longest correlation likewise occurs at \(\Gamma\) (Fig.~\ref{fig:xi_k}(d)), consistent with the zero-momentum condensate predicted by Eq.~\eqref{eq:dis}.

Close to criticality a different pattern stands out. Besides the enhancement at \(\Gamma\), the correlation lengths at the high-symmetry points
\begin{equation}
    X=(\pi/a,0),\quad
Y=(0,\pi/a),\quad
M=(\pi/a,\pi/a)
\end{equation}
grow almost as large and, more importantly, become \emph{nearly identical}, Fig.~\ref{fig:xi_k}(b).  We shall trace the near coincidence of \(\xi_{X}\), \(\xi_{Y}\), and \(\xi_{M}\) to an emergent \(\mathsf{SO(3)}\) symmetry that arises in the low-energy QED\(_3\)–CS description of the critical point.

We first review the QED$_3$ description proposed for the Laughlin–to–superfluid transition~\cite{MaissamLaughlin,song_2024_transition}.  A single boson is fractionalised into two fermions,
\begin{equation}
b(\mathbf r)=c(\mathbf r)f(\mathbf r),
\end{equation}
Our mean-field ansatz breaks the internal \(SU(2)\) gauge symmetry to its Abelian subgroup \(U(1)_a\) and leaves one emergent gauge field \(a_\mu\).  We attach the physical electric charge to \(c\) and give the partons opposite \(U(1)_a\) charges \(\mp 1\) for \(c\) and \(f\). Under our mean-field ansatz, each parton experiences \(\pi\) flux, and the combined flux matches the physical \(2\pi\) per plaquette (Fig.~\ref{fig:pd}(b)).

\begin{figure}[t]
    \centering
    \includegraphics[width=\linewidth]{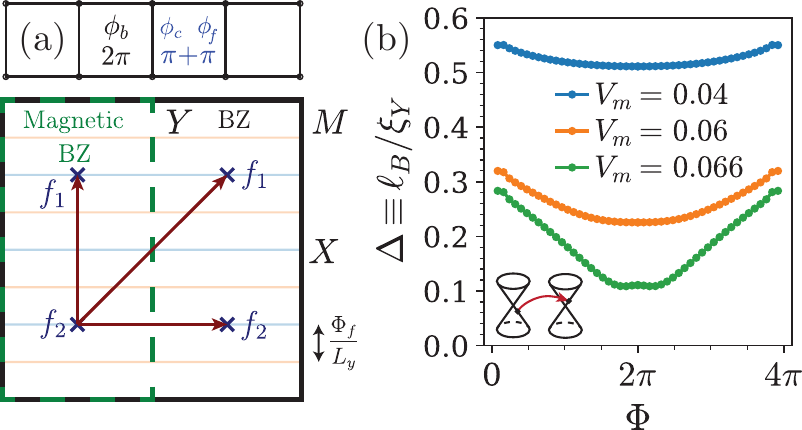}
    \caption{(a) Upper: square lattice for physical bosons and \(c\), \(f\) partons. Each boson sees \(2\pi\) flux, and each fermionic parton sees $\pi$ flux. Lower: Brillouin zone for the \(\pi\)-flux Hofstadter model of \(f\) parton. Green dashed square marks magnetic Brillouin zone.  Blue crosses are the Dirac points \(f_{1,2}\).  Orange horizontal lines label the quantized \(k_y\) values for \(f\) in the ground state on a cylinder with \(L_y=4a\).  After threading a \(\Phi_f = \pi\) flux along the cylinder the allowed momentum shift to the blue lines, crossing all Dirac points.  Red arrows indicate charge–density–wave momentum that connect the two cones.  (b) Flux–insertion spectroscopy: the inverse correlation length \(\Delta\equiv\ell_B/\xi_Y\) (interpreted as the excitation gap) versus inserted flux \(\Phi\) on a cylinder of circumference \(L_y=8a\) at \(\chi = 8000\). The inset shows the relevant particle-hole excitations.  The gap collapses near \(\Phi=2\pi\) when \(V_m\) approaches \(V_m^{c}\), revealing a massless Dirac spectrum.}
    \label{fig:xi_Dirac}
\end{figure}

If both partons occupy \(\nu = 1\) IQHE bands, the composite boson \(b = c f\) realizes the \(\nu = 1/2\) Laughlin state. Integrating out the \(c\) parton turns \(f\) into the composite fermion in the standard composite–fermion picture of bosonic Laughlin state. If \(f\) instead enters a \(C = -1\) band while \(c\) stays in \(C = +1\), the Chern–Simons terms cancel, \(a_\mu\) becomes gapless, and the boson condenses into a superfluid. We model \(f\) with a square-lattice \(\pi\)-flux Hofstadter Hamiltonian that hosts four Dirac cones at \((\pm\pi/2a,\pm\pi/2a)\) as shown in Fig.~\ref{fig:xi_Dirac}(a).  Mass terms \(m_I\) that gap these cones (equivalent to tuning a small diagonal hopping) set the Chern number of the lower band.  Integrating out \(c\) yields the two-flavour QED\(_3\)–Chern–Simons theory~\cite{MaissamLaughlin,song_2024_transition}
\begin{equation}\label{eq:qed_cs}
\mathcal L=\sum_{I=1}^{2}\bar f_I(i\slashed\partial+\slashed a-m_I)f_I+\frac{1}{4\pi}(a-A)d(a-A),
\end{equation}
with \(I=1,2\) the Dirac species per magnetic unit cell and \(a_\mu\) the emergent gauge field.  Both \(m_I>0\) reproduce the Laughlin phase, both \(m_I<0\) the superfluid, and opposite signs a Mott insulator, matching the phase diagram of the bosonic IQHE–to–Mott transition recently studied by DMRG~\cite{zeng_2020_continuous}. Different gauge choices merely shift the Dirac points and leave the charge-density-wave momentum (red arrows in Fig.~\ref{fig:xi_Dirac}) unchanged.

One may worry that a direct Laughlin–to–superfluid transition requires fine-tuning because both Dirac masses \(m_I\) must change sign simultaneously. Magnetic translation symmetry at half filling removes this concern\cite{song_2024_transition}.  With \(\pi\) flux per plaquette the parton hoppings satisfy \(T_xT_yT_x^{-1}T_y^{-1}=-1\). In the gauge we choose the translations act on the two Dirac spinors as
\begin{gather}
    T_y f_{1}=e^{ik_y}f_{2}, \quad T_y f_{2}=e^{ik_y}f_{1}\\
    T_x f_{I}=e^{ik_x+\pi I}f_{I}, \quad I=1,2
\end{gather}
Only the flavour-singlet mass \(m\sum_I\bar f_I f_I\) respects both translations, so generically \(m_1=m_2\) and a single continuous transition follows.  Because \(f_{1}\) and \(f_{2}\) are related by magnetic translations, the critical theory \eqref{eq:qed_cs} enjoys an emergent \(\mathsf{SU(2)}\) flavour symmetry, whose \(\mathbb{Z}_2\) center is already gauged and therefore yields an effective \(\mathsf{SO(3)}\).  Charge-density waves at \(X\), \(Y\), and \(M\) correspond to the triplet bilinears \(\bar f_I\sigma^i f_J\,(i=1,2,3)\). These momenta are fixed by magnetic translations and are gauge invariant even though the Dirac-cone positions move under gauge changes. The identical correlation lengths \(\xi_{X}\simeq\xi_{Y}\simeq\xi_{M} \) observed close to the critical point suggest that the three CDW operators form an \(\mathsf{SO(3)}\) vector, which provides strong evidence that the QED\(_3\)–Chern–Simons theory controls the critical behavior.

\begin{figure}[htbp]
     \centering
     \includegraphics[width=\linewidth]{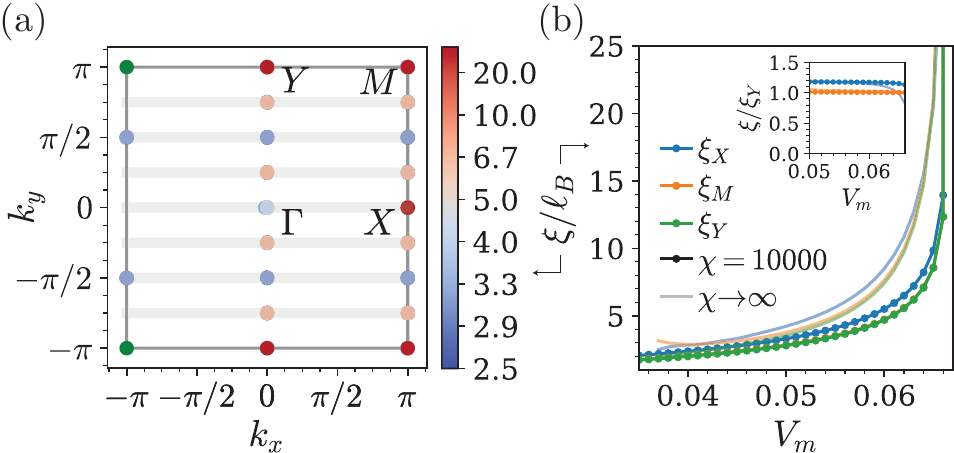}
      \caption{Neutral sector correlation lengths in the flux-threaded sector (\(\Phi=2\pi\)).  
(a) Momentum-resolved correlation lengths at the critical coupling \(V_m=0.07\). The correlation lengths at \(X\), \(Y\), and \(M\) diverge together, exceeding the cylinder circumference \(L_y=8a\).  
(b) Growth of \(\xi_X\), \(\xi_Y\), and \(\xi_M\) on approaching the transition from the Laughlin side.  
Solid curves are raw DMRG data at bond dimension \(\chi = 10000\);  
faint curves are extrapolations to \(\chi \to \infty\). The inset shows the ratio \(\xi_X/\xi_Y\) and \(\xi_M/\xi_Y\). The three correlation lengths converge toward one another and diverge with a common exponent, consistent with an emergent \(\mathsf{SO(3)}\) symmetry at criticality. }
     \label{fig:xi_exp}    
\end{figure}

\PRLsec{Revealing the critical point and the Dirac dispersion}  Quasi–1D calculations sample only the discrete transverse momentum \(k_y=2\pi n/L_y\).  In the untwisted (ground-state) sector these cuts miss the Dirac points, so the correlation lengths \(\xi_{X,Y,M}\) saturate even for very large bond dimension.  We reveal the Dirac cones by threading a physical flux \(\Phi\) through the cylinder relative to the ground state.  In the Laughlin phase a \(\Phi = 4\pi\) insertion changes the boson number by \(\Delta Q^{(b)}=2 \sigma_{xy}^{(b)}=1\).  The lowest-energy way to accommodate this extra boson is to divide the added flux equally between the two \(C=1\) partons, pumping one \(c\) and one \(f\) since \(\Delta Q^{(c/f)}= \sigma_{xy}^{(c/f)}=1\). This process leaves both IQHE gaps intact and fixes the threaded gauge flux to \(\Phi_f = \Phi/2\). The allowed \(f\)-parton momentum is therefore shifted,
\begin{equation}
    k_y^{(f)} =\frac{2 \pi}{L_y}\left(n+\frac{\Phi_f}{2 \pi}\right)
\end{equation}
The distance to the Dirac point scales linearly with \(\Phi-2\pi\), so at \(\Phi=2\pi\) the discrete \(k_y\) lines cut through all the Dirac points for even $L_y/a$ as shown in Fig.~\ref{fig:xi_Dirac}(a). 

For a Lorentz-invariant spectrum the largest correlation length sets the lowest excitation gap via \(\Delta=v_s/\xi\), where \(v_s\) is the Dirac velocity \cite{MikeDSL,transfer_matrix}. Thus \(\ell_B/\xi_Y(\Phi)\) directly probes the single-particle dispersion of the \(f\) parton. As \(V_m\) approaches \(V_m^{c}=0.07\) the gap measured collapses at \(\Phi=2\pi\) and follows the linear law 
\begin{equation}
    \Delta\propto|\Phi-2\pi|,
\end{equation}
the hallmark of a massless Dirac mode (see Fig.~\ref{fig:xi_Dirac}(b)). Very close to criticality (\(V_m=0.07\)) a tiny spontaneous \(\mathbb{Z}_2\) parity breaking appears for \(|\Phi-2\pi|\lesssim0.1\pi\), which we attribute to quasi-1D effects and analyze it in detail in the Supplementary Material~\cite{SM}. Aside from this narrow window the flux-threaded sector cleanly exposes the Dirac dispersion and provides additional evidence of the underlying QED\(_3\)–Chern–Simons theory.

Having identified the Dirac cones, we next examine how the charge-density–wave correlations behave when the \(k_y\) lines cut through the Dirac points. In the flux-threaded \(\Phi=2\pi\) sector, the charge-density–wave correlations at \(X\), \(Y\), and \(M\) now become more critical at the critical point as shown in Fig.~\ref{fig:xi_exp} (a). If we track \(\xi_X\), \(\xi_Y\), and \(\xi_M\) when we approach the critical point from the Laughlin side, the three correlation length converge toward one another and diverge with a common exponent.
\begin{equation}
    \xi_{X, Y, M} \sim\left|V_m-V_m^c\right|^{-\nu}
\end{equation}
where $\nu$ is the correlation length exponent from the singlet mass. The almost exact degeneracy of \(\xi_{X}\), \(\xi_{Y}\), and \(\xi_{M}\) strongly supports an emergent \(\mathsf{SO(3)}\) symmetry at low energies--strong evidence of the QED\(_3\)–CS theory. On the cylindrical geometry the microscopic \(C_{4}\) symmetry is absent; even if it were present it would relate only \(\xi_{X}\) and \(\xi_{Y}\), leaving \(\xi_{M}\) unconstrained. The degeneracy therefore provides direct support for the QED\(_3\)–CS description of the critical point. The precise extraction of the 2D critical exponent is subtle because once \(\xi\) surpasses \(L_y\) the system crosses over to 1D behavior. We defer that analysis to the Supplementary Material~\cite{SM}.

\PRLsec{Discussion and conclusion} We have demonstrated a direct, continuous transition from the bosonic \(\nu=1/2\) Laughlin state to a zero-momentum superfluid.  
Flux insertion collapses the transfer-matrix gap \emph{linearly} with enclosed flux, providing a clear dynamical signature of massless Dirac quasiparticles at the critical point.  
The momentum-resolved correlation lengths show that the charge density wave (CDW) operators at the lattice momenta \(X\), \(Y\), and \(M\) diverge with the same exponent, revealing an emergent \(\mathsf{SO(3)}\) symmetry. 
 Taken together, these complementary signatures provide the first microscopic evidence that the transition is governed by the long-conjectured QED\(_3\)–Chern–Simons (CS) fixed point. Our results also provide the first numerical evidence that the two-flavor level-\(1\) QED\(_3\)--CS theory is stable against gauge field fluctuations, which has only been inferred from large-\(N\) analysis \cite{MaissamLaughlin,JYL}.

Our numerics itself cannot yet exclude a hidden topological order on the superfluid side.  However, the observed CDW instability argues strongly against this possibility.  Neither the $\nu=1/2$ Laughlin liquid nor a uniform superfluid favors CDW order, yet at the critical point all three symmetry-inequivalent CDW operators become sharply enhanced. Such behavior typifies a confinement transition governed by the QED–CS theory, beyond which the resulting superfluid is topologically trivial.

The triangular-lattice results in the Supplementary Material~\cite{SM} show that neither the lattice details, nor the boundary orientation, nor the exact position of the folded CDW vector affects the universal physics of the transition.  A single continuous Laughlin–to–superfluid change occurs, the low-energy spectrum is Dirac-like, and the three CDW operators form an \(SO(3)\) triplet with identical scaling.  These findings reinforce the claim that the QED\(_3\)–CS theory is a universal description of the transition.

Transitions described by the QED\(_3\)–CS theory lie beyond Landau's paradigm and are now being explored in moiré platforms including twisted homobilayer transition-metal dichalcogenides and rhombohedral pentalayer graphene.  In Supplementary Material we discussed a similar bosonic FQH-SF transition at $\nu=3/4$ and relevance to moiré systems.The flux-threading spectroscopy and momentum-resolved diagnostics introduced here can be applied directly to those materials. 

\PRLsec{Note added} During writing of this manuscript, we become aware of Ref.~\onlinecite{ZiYangLaughlin,ZiYangLaughlin2,Lotric2025-mt}, which studies a similar transition in lattice hopping systems.

\textbf{Acknowledgments.} We thank Yin-Chen He, Maissam Barkeshli, and Zi-Yang Meng for useful discussions. This work is partially supported by the Simons Collaboration on Ultra-Quantum Matter, which is a grant from the Simons Foundation
(Grant No. 651446, TS, and Grant No. 1151944, MZ). T.W. is supported by the  Simons Collaboration on Ultra-Quantum Matter via M.Z., and the Heising-Simons Foundation, the Simons Foundation, and NSF grant No. PHY-2309135 to the Kavli Institute for Theoretical Physics (KITP). T.S. is also supported by NSF grant DMR-2206305. M.Z. is also supported by the U.S. Department of Energy, Office of Science, Office of Basic Energy Sciences, Materials Sciences and Engineering Division under Contract No. DE-AC02-05-CH11231 (Theory of Materials program KC2301). X.Y.S. is supported by the Gordon and Betty Moore Foundation EPiQS Initiative through Grant No. GBMF8684 at the Massachusetts Institute of Technology, and by Early Career Scheme of Hong Kong Research Grant Council with grant No. 26309524. This research uses the Lawrencium computational cluster provided by the Lawrence Berkeley National Laboratory (Supported by the U.S. Department of Energy, Office of Basic Energy Sciences under Contract No. DE-AC02-05-CH11231).

\input{arxiv.bbl}
\newpage

\onecolumngrid

\vspace{0.3cm}

\supplementarysection

\newpage

\input{supp_arxiv.tex}

\vfill 

\end{document}

%% file: supp_arxiv.tex
\begin{center}
\Large{\bf Supplemental Material for ``Emergent QED$_3$ at the bosonic Laughlin state to superfluid transition''}
\end{center}

\section{Hamiltonian and DMRG setup}

We work on an infinite cylinder with circumference \(L_y\).  In Landau gauge \(\mathbf A=(0, Bx)\) the lowest-Landau-level orbitals are  
\begin{equation}
\varphi_n(x,y)=\frac{\exp\left[i k_y y-\left(x-k_y\ell_B^{2}\right)^{2}/(2\ell_B^{2})\right]}
                     {\sqrt{L_y \ell_B \sqrt{\pi}}},
\qquad
k_y=\frac{2\pi n}{L_y},
\quad n\in\mathbb Z .
\end{equation}
Each orbital is treated as a one-dimensional site in the matrix-product-state ansatz.  The on-site Hilbert space is truncated to occupations \(n_n \leq 3\).  Increasing the cutoff to \(n_n=4\) changes the ground-state energy density by less than \(5\times10^{-4}V_0\), so the choice \(n_n\le3\) is sufficient even in the superfluid regime, where spontaneous \(U(1)\) breaking is absent on a quasi-one-dimensional cylinder.

Before projection to the LLL the Hamiltonian is  
\begin{equation}
H=\frac12\int_{\mathbf r,\mathbf r^{\prime}}\hat\rho_{\mathbf r}\delta_{\mathbf r-\mathbf r^{\prime}}\hat\rho_{\mathbf r^{\prime}}
   -2 V_m\sum_{j=1}^{2}\int_{\mathbf r}\hat\rho_{\mathbf r}\cos\left(\mathbf G_j\cdot\mathbf r\right) .
\end{equation}
The first term is the Haldane \(V_0\) pseudopotential corresponding to a contact interaction.  The second term is a periodic potential with one flux quantum per real-space unit cell.  For the square lattice used in the main text  
\begin{equation}
\mathbf G_{1,2}=(\pm 2\pi/a,0),\ (0,\pm 2\pi/a),
\quad
a=\sqrt{2\pi}\ell_B .
\end{equation}
Appendix~\ref{sec:tri} presents analogous \(\mathbf G_j\) for a triangular lattice that also encloses total flux \(2\pi\) per unit cell.

The lattice reduces continuous translation around the cylinder to discrete shifts by one lattice spacing \(a\).  With \(m=L_y/a\) unit cells the allowed crystal momentum are  
\begin{equation}
k_y=\frac{2\pi n}{L_y},
\qquad
n=0,\dots,m-1 ,
\end{equation}
so the computation is more expensive than in a homogeneous fractional-quantum-Hall cylinder due to the reduced symmetry.

We perform infinite DMRG with bond dimension up to \(\chi=15000\).  The four-index interaction tensor is compressed by discarding matrix elements with \(|k_i-k_j|>8\) and singular values below \(10^{-8}\); the resulting MPO bond dimension does not exceed \(1200\). Energies converge to \(10^{-8}V_0\) per flux quantum, and correlation lengths change by less than two percent between the two largest \(\chi\) values.

To characterize superfluidity we evaluate the real-space Green function  
\begin{equation}
G(x)=\left\langle b^{\dagger}(x,0)b(0,0)\right\rangle .
\end{equation}
First we compute the orbital correlation matrix 
\(C_{ij}=\langle b_{i}^{\dagger} b_{j}\rangle\) 
directly from the ground-state MPS, where the site operator is related to the continuum field by  
\(b(x,y)=\sum_{j}\varphi_{j}(x,y)b_{j}\).  
Using the known orbitals $\varphi_{j}$ we convert \(C_{ij}\) to \(G(x)\).  
Algebraic decay of \(G(x)\) signals the superfluid phase, while exponential decay appears in the Laughlin phase. To characterize the Laughlin phase, we extract the orbital entanglement spectrum (OES) by cutting the infinite MPS at a chosen bond.  Spectra extracted at the two inequivalent bonds are nearly identical; one representative cut is shown in the main text.

Momentum-resolved correlation lengths are extracted from the eigenvalues of the infinite-MPS transfer matrix.  
We choose a unit cell of length \(L_x\) along \(\hat x\); the transfer matrix \(T\) advances the state by one cell and factorises into blocks labelled by charge \(Q\) and conserved momentum \(k_y\).  
For each \(k_y\) we solve  
\begin{equation}
T(k_y)\,v_n=\lambda_n\,v_n,
\end{equation}
ordering the eigenvalues by magnitude.  In the \((Q=0,k_y=0)\) block the leading value is \(\lambda_0=1\), the identity.  
Any sub-leading eigenvalue can be written  
\begin{equation}
\lambda_n=\exp \bigl[-L_x/\xi_n+i\,k_x^{(n)}L_x\bigr],
\label{eq:lambdadef_app}
\end{equation}
so the correlation length and momentum follow directly:  
\begin{equation}
\xi_n=\frac{L_x}{-\ln|\lambda_n|},\qquad
k_x^{(n)}=\frac{\arg\lambda_n}{L_x}.
\end{equation}
Collecting the first few eigenvalues from every \(k_y\) block yields the triplets \((k_x,k_y,\xi_n)\) that form the map displayed in the main text.  

To remove finite bond-dimension effects, we track a single excitation branch as \(\chi\) grows.  For each \(\chi\) we retain the twenty largest eigenvalues in the neutral sector and convert them to inverse lengths \(\epsilon_i=L_x/\xi_i\) using Eq.~\eqref{eq:lambdadef_app}. The two smallest inverse lengths \(\epsilon_1 \leq \epsilon_2\) with closest momentum \(k_x^{(i)}\) are taken to lie on the same branch. We denote the level spacing \(\delta=\epsilon_2-\epsilon_1\).  As \(\chi\to\infty\) the spacing \(\delta\) vanishes, and \(\epsilon_1\) approaches its asymptotic value \(\epsilon_\infty\). We therefore fit the four largest-\(\chi = 4800,6400,8000,10000\) points to fit the scaling hypothesis,
\begin{equation}
\epsilon_1(\delta)=\epsilon_\infty+A\,\delta^{p},\qquad p\approx1,
\end{equation}
and the extrapolated length is given by \(\xi_\infty=1/\epsilon_\infty\).  The resulting values are plotted as translucent curves in Fig.~\ref{fig:xi_exp}(b). The consistency of the fit confirms that the bond dimension employed already places the calculation in the universal scaling regime.

Three numerical ingredients proved essential in this work: (i) flux threading shifts the discrete quasi-1D momentum and reveals the Dirac dispersion; (ii) momentum-resolved correlation lengths reveal the emergent \(\mathsf{SO(3)}\) symmetry; and (iii) modern MPO compression enables bond dimensions sufficient to capture a gapless superfluid on the cylinder.  Together, these advances turn quasi-1D DMRG into a precise probe of 2D quantum criticality and open the door to systematic studies of QED\(_3\) and Chern–Simons critical points across a wide range of strongly correlated lattice models.

\section{Single–particle dispersion in a projected periodic potential}

Because a single flux quantum threads each unit cell, the magnetic translations generated by the real-space primitives commute, \(T(\mathbf a_1)T(\mathbf a_2)=T(\mathbf a_2)T(\mathbf a_1)\).  Hence Bloch states are simultaneous eigenstates,
\begin{equation}
T(\mathbf a_j)|\mathbf k\rangle = e^{i\mathbf k\cdot\mathbf a_j}|\mathbf k\rangle,
\qquad j=1,2 .
\end{equation}

A monochromatic modulation of strength \(V_m\) is written as \(\widehat V = -2V_m\sum_j \cos(\mathbf G_j\cdot\mathbf r)\).  Projecting the plane waves to the lowest Landau level multiplies them by the Gaussian form factor and yields the guiding-centre Hamiltonian
\begin{equation}
H_{\mathrm{pot}} = 2V_m\sum_j e^{-\mathbf G_j^{2}\ell_B^{2}/4}
                   \cos(\mathbf G_j\cdot\mathbf R),
\end{equation}
where \(\rho_{\mathbf G}\equiv e^{i\mathbf G\cdot\mathbf R}\) and \(\mathbf R\) denotes the guiding-center coordinates.

The identity \(\rho_{\mathbf q}=T(-\ell_B^{2}\hat{\mathbf z}\times\mathbf q)\) together with the flux-quantum condition \(\mathbf a_1\times\mathbf a_2=2\pi\ell_B^{2}\) gives the dual vectors
\begin{equation}
    \mathbf b_1=-\mathbf a_2, \quad \mathbf b_2=\mathbf a_1
\end{equation}
and, for a triangular lattice a third vector \(\mathbf b_3=\mathbf a_1+\mathbf a_2\) appears.  Each \(\rho_{\mathbf G_j}=T(-\mathbf b_j)\) commutes with all lattice translations and therefore acts diagonally, \(\rho_{\mathbf G_j}|\mathbf k\rangle=e^{i\mathbf k\cdot\mathbf b_j}|\mathbf k\rangle\).  The dispersion is
\begin{equation}
\varepsilon'(\mathbf k)=
   2V_m e^{-\mathbf G^{2}\ell_B^{2}/4}
   \sum_j \cos(\mathbf k\cdot\mathbf b_j)
   =2V_m e^{-\mathbf G^{2}\ell_B^{2}/4}
   \sum_j \cos(\mathbf k\cdot\mathbf a_j),
\end{equation}
the second form following from \(\mathbf b_1=-\mathbf a_2\), \(\mathbf b_2=\mathbf a_1\). However, we adopt a gauge in which $\tilde{T}_j = -T_j$, so that the point-group operations act trivially. This choice amounts to shifting $\mathbf{k} \mapsto \mathbf{k} + (\pi,\pi)$, producing the dispersion shown in the main text,
\begin{equation}
\varepsilon(\mathbf{k}) = -2V_m e^{-\mathbf{G}^{2}\ell_B^{2}/4}
\sum_j \cos\bigl(\mathbf{k}\!\cdot\!\mathbf{a}_j\bigr).
\end{equation}
The minimum occurs at $\mathbf{k}=0$, so a sufficiently large $V_m$ drives Bose–Einstein condensation into the zero-momentum state.

\section{Circumference dependence of the transition}

\begin{figure}[t]
    \centering
    \includegraphics[width=\linewidth]{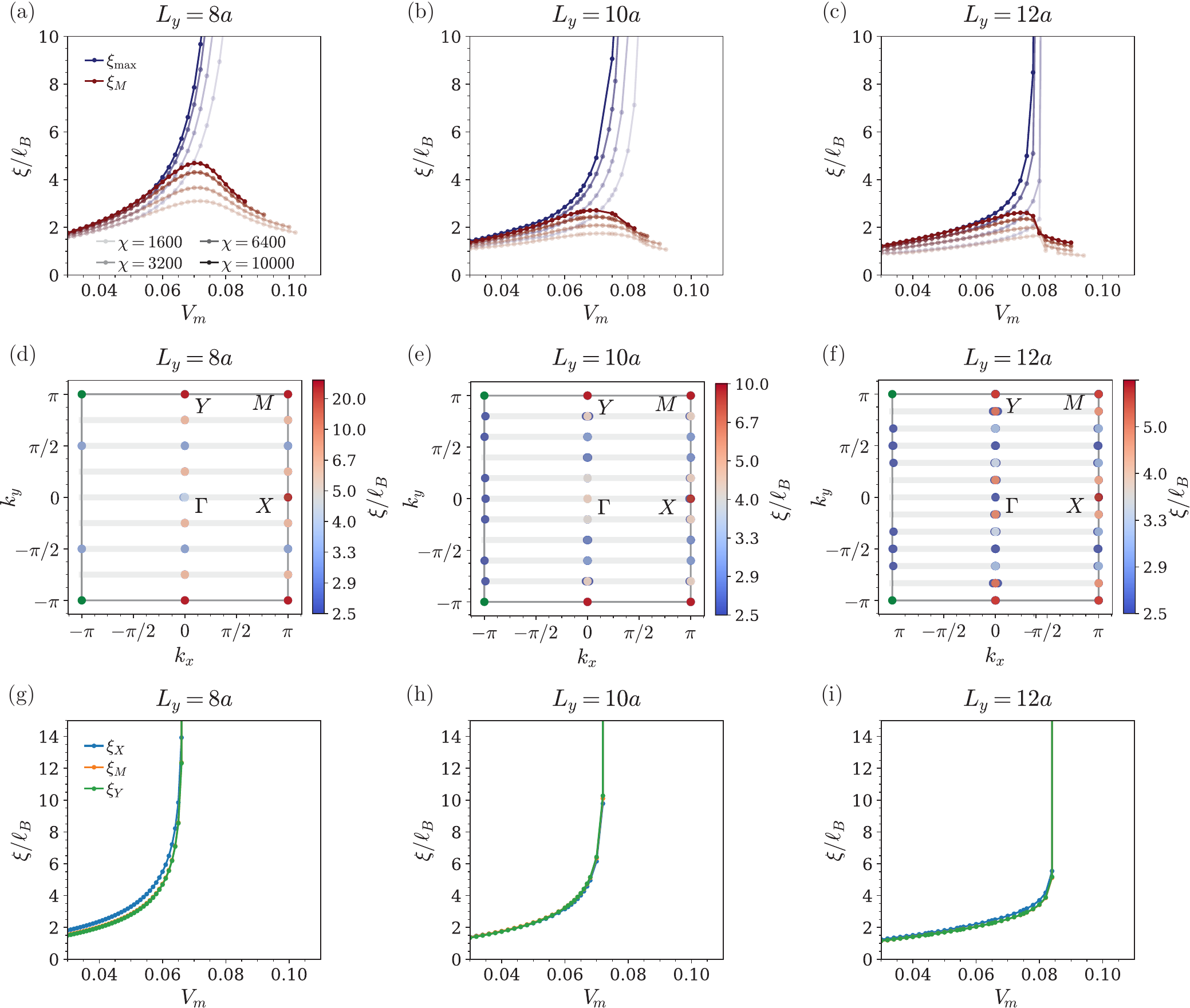}
    \caption{Correlation-length diagnostics for cylinders with circumference \(L_y/a = 8, 10, 12\).  
(a)-(c)~Ground-state neutral sector: longest correlation length \(\xi_{\max}\) and \(M\)-point correlation length \(\xi_M\) versus potential strength \(V_m\).  
Their contrasting trends locate a single Laughlin–to–superfluid transition whose critical coupling shifts slightly with \(L_y\).  
(d)-(f)~Flux-threaded neutral sector with \(\Phi = 2\pi\): momentum-resolved correlation lengths at the respective critical couplings \(V_m = 0.070, 0.072,\) and \(0.084\) at bond dimension \(\chi = 10000\).  
The color scale highlights converging peaks at \(X\), \(Y\), and \(M\), evidencing an emergent degeneracy.  
(g)-(i)~Correlation lengths at the three CDW momenta for the flux-threaded sector.  
The simultaneous divergence of \(\xi_X\), \(\xi_Y\), and \(\xi_M\) confirms an emergent \(\mathsf{SO(3)}\) symmetry at all circumferences.}
    \label{fig:L}
\end{figure}

A natural consistency test for quasi-1D simulations is to check that the qualitative physics persists as the cylinder is widened.  Fig.~\ref{fig:L} therefore repeats our correlation-length analysis for three circumferences, \(L_y/a = 8, 10,\) and \(12\).

Fig.~\ref{fig:L}(a)–(c) show the ground-state sector \((\Phi = 0)\).  For every width we recover the same phase diagram: a fully gapped Laughlin liquid at weak potential, and a gapless superfluid at strong potential. At a single intermediate coupling, the \(M\)-point length \(\xi_M\) peaks while the longest neutral length \(\xi_{\max}\) rises sharply.  The inferred critical coupling shifts only slightly from \(L_y=8a\) to \(L_y=12a\), showing a modest finite-size drift rather than a change of universal physics.

To highlight universal features we turn to the flux-threaded sector with \(\Phi = 2\pi\), which forces the discrete transverse momenta through the Dirac cones.  The momentum-resolved maps in panels~\ref{fig:L}(d)–(f) are nearly identical for all circumferences. Correlation-length peaks appear simultaneously at the three charge-density-wave wave-vectors \(X\), \(Y\), and \(M\).  Following those peaks as \(V_m\) is tuned, Fig.~\ref{fig:L}(g)–(i) show that \(\xi_X\), \(\xi_Y\), and \(\xi_M\) collapse onto a common curve and diverge together, confirming the emergent \(\mathsf{SO(3)}\) symmetry predicted by the QED\(_3\)–Chern–Simons theory.

In the ground state sector, wider cylinders place the \(k_y\) grid of the fermionic parton \(f\) closer to the Dirac points, but they also carry more entanglement.  The larger bond dimension required for full convergence is not yet reached for \(L_y/a = 10\) and \(12\), so the reduced growth of \(\xi_{\max}\) in the ground-state data reflects limited \(\chi\) rather than intrinsic physics.

We also simulated \(L_y/a = 6\) (not shown).  It exhibits the same qualitative behavior at the criticality, but its critical coupling is higher, \(V_m^{c} \approx 0.12\). Smaller system sizes generally require a stronger potential to stabilize long-range order, so this shift is another finite-size effect that complicates comparison with wider systems; we omit those data for clarity.

All accessible circumferences therefore support a single continuous Laughlin–to–superfluid transition with identical symmetry-enlargement signatures, reinforcing the robustness of the QED\(_3\)–Chern–Simons critical point.

\section{Interpretation of the flux-threaded sector}

\begin{figure}[htbp]
     \centering
     \includegraphics[width=0.45\linewidth]{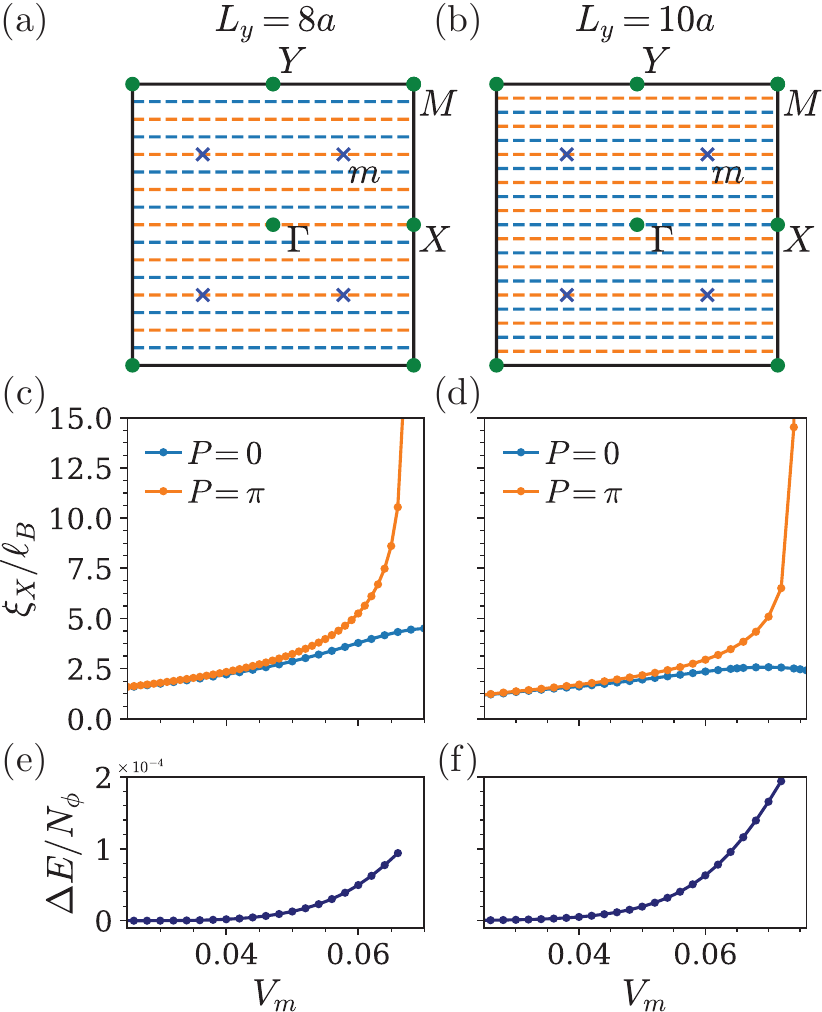}
     \caption{
Panels (a) and (b): discrete \(k_y\) lines (dashed) sampled by the parton \(f\) on cylinders with circumference \(L_y/a = 8\) and \(10\).  
Blue lines correspond to the untwisted sector \(\Phi = 0\); orange lines correspond to the flux–threaded sector \(\Phi = 2\pi\), which forces the cut through every Dirac point (blue crosses).  High–symmetry momentum are marked by green dots.  
Panels (c) and (d): evolution of the \(X\)-point correlation length \(\xi_X\) in the two sectors.  
In the flux–threaded case (\(\Phi = 2\pi\), orange) \(\xi_X\) diverges on approaching \(V_m^{c}\), whereas in the untwisted sector (\(\Phi = 0\), blue) it remains finite.  
Panels (e) and (f): energy density difference between the two sectors; the cost is negligible deep in the Laughlin phase and grows smoothly as the transition is approached.}
     \label{fig:xi_cut}    
\end{figure}

Quasi-1D simulations quantize the transverse momentum as \(k_y = 2\pi n/L_y\).  In the untwisted ground-state sector these discrete cuts usually miss the Dirac points, so the correlation lengths at \(X\), \(Y\), and \(M\) saturate even at very large bond dimension. We expose the Dirac cones by threading a uniform flux \(\Phi\) through the cylinder, which shifts the allowed parton momenta.

The charge response obeys the Streda formula. 
In the bosonic Laughlin phase the Hall conductance is \(\sigma_{xy}^{(b)} = 1/2\). Inserting \(\Phi = 4\pi\) therefore creates one extra boson,
\begin{equation}
\Delta n_b = \frac{\sigma_{xy}^{(b)}\Phi}{2\pi}=1 .
\end{equation}
This composite boson splits into one \(c\) and one \(f\) parton, so 
\begin{equation}
    \Delta n_c = \Delta n_f = \Delta n_b = 1
\end{equation}
In the bosonic Laughlin phase each parton sits in a \(C = +1\) IQHE band with \(\sigma_{xy}^{(c)} = \sigma_{xy}^{(f)} = 1\). Pumping one particle of either species needs \(2\pi\) of flux, precisely half of \(\Phi\) since \(\sigma_{xy}\) doubles.  The lowest-energy configuration therefore divides the physical flux equally,
\begin{equation}
\Phi_c = \Phi_f = \frac{\Phi}{2},
\end{equation}
leaving both IQHE gaps intact.  The \(f\) momentum grid is shifted by
\begin{equation}
k_y^{(f)} = \frac{2\pi}{L_y} \left(n+\frac{\Phi_f}{2\pi}\right),
\end{equation}
so for even \(L_y/a\) every Dirac point is crossed when \(\Phi = 2\pi\).  Fig.~\ref{fig:xi_cut} (a,b) illustrate the resulting cuts for \(L_y/a = 8\) and \(10\).

Fig.~\ref{fig:xi_cut} (c,d) compare the \(X\)-point length \(\xi_X\) in the untwisted sector \((\Phi = 0)\) and in the flux-threaded sector \((\Phi = 2\pi)\).  With \(\Phi = 0\) the cuts avoid the cones and \(\xi_X\) stays finite. In contrast, with \(\Phi = 2\pi\) they hit the cones and \(\xi_X\) diverges as \(V_m\) approaches \(V_m^{c}\), revealing the gapless Dirac dispersion. The energy–density difference between the untwisted and flux–threaded sectors \(\Delta E=(E_{\Phi=2\pi}-E_{\Phi=0})/N_\phi\) is shown in Fig.~\ref{fig:xi_cut} (e,f).  Deep inside the Laughlin phase the splitting is exponentially small \(\Delta E\propto e^{-L_y/\xi}\) because the \(\Phi=2\pi\) sector represents a topologically degenerate ground state.  Closer to the transition the Dirac cones dominate the low-energy spectrum and the scaling crosses over to an algebraic form \(\Delta E\propto\ell_B/L_y\), consistent with a linearly dispersing critical mode.

The same construction fails in the superfluid phase, where the parton Chern numbers change to \(C_c = +1\) and \(C_f = -1\). The net flux required to pump one \(c\) and one \(f\) parton is therefore zero, so no fixed partition of the \(4\pi\) physical flux minimizes the energy. DMRG consequently struggles to converge and we do not analyze the flux-threaded data there.

Flux threading thus offers a controlled route to force the parton momenta through the Dirac cones, extract the linear dispersion, and sharpen the critical behavior without increasing the cylinder width.  The technique works throughout the Laughlin phase and right up to the critical point.

\section{\(\mathbb{Z}_2\) parity breaking close to \(\Phi = 2\pi\)}

\begin{figure}[htbp]
    \centering
    \includegraphics[width=0.4\linewidth]{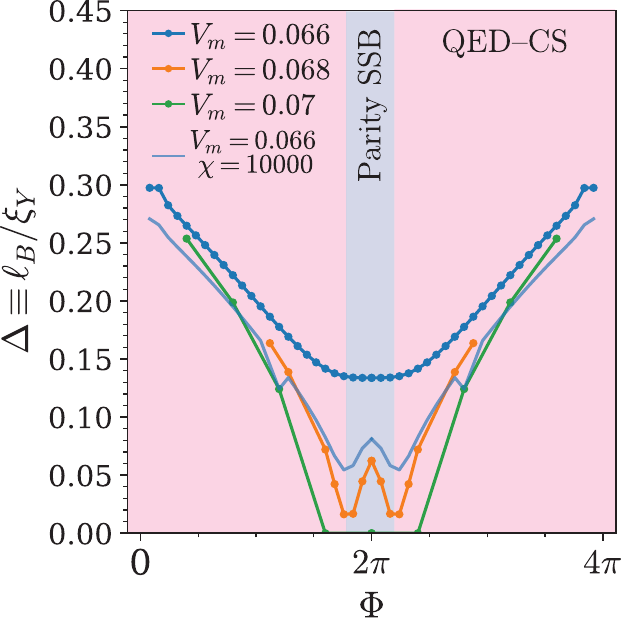}
    \caption{Flux–insertion spectroscopy on the \(L_y=8a\) cylinder close to criticality.  
The plotted quantity \(\Delta\equiv\ell_B/\xi_Y\) is the inverse \(Y\)-point correlation length and may be interpreted as the lowest excitation gap.  
Solid curves are \(\chi=6400\) data; the translucent curve shows data at \(\chi=10000\) for \(V_m=0.066\).  
Just below criticality (\(V_m = 0.066, 0.068\)) the curves develop two dips at $\Phi = 2\pi \pm 0.1 \pi$, and the gap remains nonzero at $\phi = 2\pi$. We interpret $\Phi = 2\pi \pm 0.1 \pi$ as a weak parity-breaking phase due to finite-size effect.}

    \label{fig:parity}
\end{figure}

Fig.~\ref{fig:parity} displays the flux-insertion spectrum \(\Delta=\ell_{B}/\xi_{Y}\) on an \(L_{y}=8a\) cylinder for several couplings near the transition. Deep in the Laughlin phase \((V_{m} \lesssim 0.06)\) the data follow the massive Dirac form
\begin{equation}
    \Delta^{2}=m^{2}+v_{s}^{2}(\Phi-2\pi)^{2}
\end{equation}
with a single minimum at \(\Phi=2\pi\).  Closer to criticality \((V_{m}=0.066\), the minimum splits into two shallow dips at \(\Phi=2\pi\pm0.1\pi\) at sufficiently large bond dimension \(\chi=10000\), while the gap at \(\Phi=2\pi\) remains finite. The same splitting is visible already at \(\chi=8000\) for \(V_{m}=0.068\); and at the transition \(V_{m}^{c}=0.070\) the gap collapses in the entire range of \(1.9\pi<\Phi<2.1\pi\).

We interpret the double minimum as a quasi-1D instability that breaks the parity symmetry.  Once the dynamical \(U(1)\) gauge field is integrated out on the cylinder a single Tomonaga–Luttinger liquid with central charge \(c=1\) remains.  In one dimension four-fermion interactions become marginal. A negative coupling in principle can flow to strong coupling and generates an Ising mass  
\begin{equation}
    m_{z}\propto\bar{f}\sigma^{3}f ,
\end{equation}
gapping out both Dirac cones. The shaded band in Fig.~\ref{fig:parity} indicates this parity-broken window.  We thus regard the effect as a finite-size effect specific to the quasi-1D geometry rather than a genuine instability of the \(2+1\)-dimensional QED\(_3\)–Chern–Simons fixed point.

\section{Lattice effects on the transition} \label{sec:tri}

\begin{figure}[htbp]
    \centering
    \includegraphics[width=\linewidth]{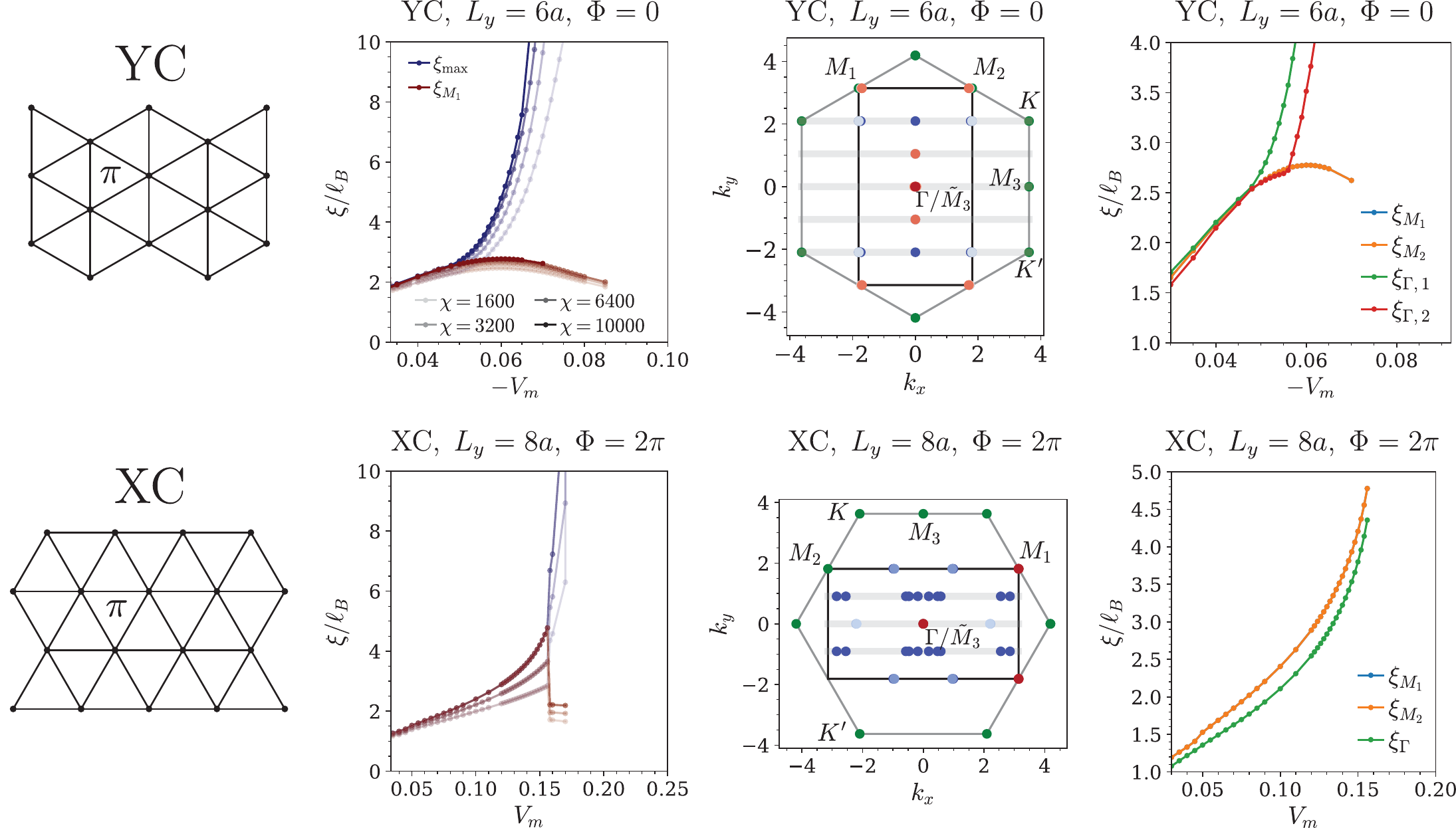}
    \caption{Triangular–lattice diagnostics on two cylinder geometries.  
Panels (a)–(d) show the triangular lattice with YC boundary condition with circumference \(L_y = 6a\) in the ground-state sector \((\Phi = 0)\); panels (e)–(h) show the XC boundary condition with \(L_y = 8a\) in the flux-threaded sector \((\Phi = 2\pi)\).  
(a,e) Real-space YC and XC triangular lattices, each pierced by a uniform \(\pi\) flux per triangular plaquette.  
(b,f) Longest neutral correlation length \(\xi_{\max}\) and \(M_1\)-point length \(\xi_{M_1}\) versus potential strength \(V_m\).  Their simultaneous upturn identifies a single Laughlin–to–superfluid transition whose critical coupling differs between YC and XC.  
(c,g) Momentum-resolved neutral lengths at the respective critical couplings, \(V_m = 0.055\) (YC) and \(0.156\) (XC).  The Brillouin zone folds differently on the two cuts: for both, the momentum \(M_3\) folds onto \(\Gamma\).  Peaks at the inequivalent CDW momenta \(M_{1,2}\) and \(\Gamma\) reproduce the pattern seen on the square lattice.  
(d,h) Growth of the CDW correlation lengths on approaching the transition.  In the flux-threaded sector (h) the \(\Gamma\) branch tracks the \(M_3\) CDW, whereas in the ground-state sector (d) the same branch is quickly dominated by the Goldstone mode because the \(k_y\) cuts miss the Dirac point.  In both geometries the joint divergence of \(\xi_{M_1}\), \(\xi_{M_2}\), and \(\xi_{\Gamma}\) confirms an emergent \(\mathsf{SO(3)}\) symmetry.}
    \label{fig:triangle}
\end{figure}

A decisive universality test is to place the bosons on a triangular lattice. We take Bravais vectors \(\mathbf a_1=(a,0)\) and \(\mathbf a_2=(a/2,\sqrt3\,a/2)\) and pierce every plaquette with a uniform \(\pi\) flux so that a single flux quantum threads the doubled real-space cell used on a cylinder.  The three shortest reciprocal vectors are \(\mathbf G_{1,2}=(\pm2\pi/a,0)\) and \(\mathbf G_3=-\mathbf G_1-\mathbf G_2\).  After projection to the LLL the potential retains the form
\begin{equation}
H_{\text{pot}}=-2V_m\sum_{j=1}^{3}\int_{\mathbf r} \rho_{\mathbf r}\cos(\mathbf G_j \cdot \mathbf r),
\end{equation}
Because the bosons are still at half filling, the \(f\) parton still hosts two Dirac cones $I=1,2$. On the triangular lattice the magnetic-translation algebra pins the gauge-invariant density waves to the three mid-reciprocal vectors
\begin{equation}
M_{1,2,3}=\tfrac12\mathbf G_{1,2,3},
\end{equation}
which play the same role as the $X, Y$, and $M$ points did on the square lattice.

To impose cylindrical boundary conditions we identify lattice sites separated by a chosen primitive vector, rendering that direction periodic with circumference \(L_y\) while the perpendicular direction remains open.  On the triangular lattice two orientations are standard.  As illustrated in Fig.~\ref{fig:triangle}(a,e), a YC cylinder glues sites displaced by multiples of \(\mathbf a_{2}\), whereas an XC cylinder glues sites displaced by multiples of \(\mathbf a_{1}+\mathbf a_{2}\). The resulting wrapping doubles the unit cell in the direction perpendicular to the bonds and folds the two-dimensional Brillouin zone into a rectangle, shown in Fig.~\ref{fig:triangle}(c,g).  Under this folding the CDW vector \(M_{3}\) lands at the zone center \(\Gamma\), while \(M_{1}\) and \(M_{2}\) map to the zone corners. Each boundary condition therefore makes two CDW wave-vectors explicit and hides the third inside the leading \(\Gamma\) branch.

We first consider the YC cylinder in the untwisted ground-state sector \(\Phi=0\). As shown in Fig.~\ref{fig:triangle} (b), the longest neutral correlation length \(\xi_{\max}\) grows sharply and appears to diverge at \(V_m^{c}\simeq0.055\), while the CDW length \(\xi_{M_1}\) exhibits a pronounced peak at the same coupling, reproducing the behavior on the square lattice.  The momentum-resolved map at \(V_m^{c}\) shows two equal maxima at \(M_{1}\) and \(M_{2}\) in Fig.~\ref{fig:triangle} (c). The leading branch at \(\Gamma\) represents the folded \(M_{3}\) operator. One of the correlation lengths collapses onto the same divergent curve as \(\xi_{M_{1}}\) and \(\xi_{M_{2}}\) when \(V_m\to V_m^{c}\) as shown in Fig.~\ref{fig:triangle} (d).  On the superfluid side the branch at \(\Gamma\) is eventually dominated by the Goldstone mode. The three-fold degeneracy at criticality provides the anticipated \(\mathrm{SO}(3)\) fingerprint of the QED\(_3\)-CS theory. Interestingly, once the transition is crossed the condensate momentum lies at \(M_{1,2}\) rather than at the single-particle minimum \(\Gamma\), which we defer to future work for further exploration.

The XC cylinder yields the same universal picture. Similar to the square lattice case, the Dirac nodes can be reached by threading \(\Phi=2\pi\), which shifts every discrete momentum by \(\Delta k_y=\Phi/(2L_y)\).  In this flux-threaded sector both the longest neutral length \(\xi_{\max}\) and the CDW length \(\xi_{M_1}\) diverge at \(V_m^{c}\simeq0.156\).  The momentum map recorded at the critical coupling displays three equal peaks at \(M_{1}\), \(M_{2}\), and \(\Gamma\). Here \(\xi_{\Gamma}\) corresponds to the unfolded \(M_{3}\) CDW opertor, whose correlation length overtakes the Goldstone branch because the \(k_y\) grid now passes directly through the Dirac points.  Thus the \(M_{1},M_{2},M_{3}\) triplet again becomes degenerate at criticality, confirming that the emergent \(\mathrm{SO}(3)\) symmetry persists in the XC geometry.

\section{Bosonic Fractional quantum Hall(FQH) to superfluid transition at $\nu=-3/4$ and connection to Moire systems}

In this section we discuss a FQH-SF transition theory at filling $\nu=3/4$ closely related to the transition studied in the main text. This theory is also connected to the putative continuous transition from composite Fermi liquid(CFL) to Fermi liquid(FL) at filling $\nu=-3/4$, potentially realizable in Moire TMD bilayer systems, at filling $\nu=-3/4$.

We start from a parton decomposition for the electrons $c_s=bf_s$ where $b,f_s$ are charged holons and spinons, respectively and a $U(1)$ gauge field $a$ that couples oppositely to $b,f_s$.This construction comes with the gauge constraint on the particle numbers $n_c=n_b=n_f$. To describe a composite Fermi liquid of electrons at filling $\nu=-3/4$, ref \cite{song_2024_transition,dong2023_composite} proposed to think of the system as a composite of a $\nu=-1$ IQH state and $\nu=1/4$ composite Fermi liquid state. The transition to neighboring Fermi liquid is described by the transition of the $\nu=1/4$ system, from composite Fermi liquid to Fermi liquid. The resulting Fermi liquid hence has a large Hall conductivity from the $\nu=-1$ IQH sector. This is controlled by the holon transition from FQH at $\nu=1/4$ to superfluid, with the spinons filling a Fermi surface, e.g.
\begin{align}
    \mathcal L=\mathcal L_{FQH-SF}[b;a]+\mathcal L_{f_s;a},\nonumber\\
    \mathcal L_{f_s;a}=\sum_s f_s^\dagger(i\partial_t-\mu) f_s+\sum_{\mu=x,y}\frac{|(i\partial_\mu-a_\mu)f_s|^2}{2m},
\end{align}
where $\mathcal L_{FQH-SF}[b;a]$ describes the holon sector transition from ref \cite{song_2024_transition} and $\mathcal L_{f_s;a}$ describes the spinon Fermi surface state.

Here we discuss a different route: We consider the holon FQH state at $\nu=3/4$ and spinons fill a Fermi sea at $\nu=3/4$, the physical system of which is a particle-hole conjugate of $\nu=-3/4$ composite Fermi liquid as relevant to twisted MoTe$_2$. To describe the holon FQH state, we further preform a parton for the holon $b=f_1f_2$ where $f_{1,2}$ are fermionic particles, both at the same filling $3/4$ as required by gauge constraint. $f_{1,2}$ each fill Chern bands with total Chern number $C_1=1,C_2=3$ which constitutes the bosonic Jain state of $\nu=p/(p+1)$ for $p=3$. When the Chern band for $f_2$ goes through a transition to Chern number $C_2=-1$, the bosons form a superfluid. Hence the system enters the Fermi liquid phase. Note that the Chern bands can be formed at mean-field level when $f_{1,2}$ each sees $\pm 1/4$ flux quantum per unit cell, hence quadrupling the unit cell and the corresponding effective filling is $3$. This allows the $f_i$'s to completely fill Chern bands.

The critical theory is controlled by the $\nu=3/4$ boson FQH to superfluid transition,
\begin{equation}
    \mathcal L_{b;a,\alpha}=\sum_{i=1}^4 \bar\psi_i i\slashed\partial_\alpha \psi_i+m\sum_i \bar\psi_i\psi_i+\frac{2}{4\pi}\alpha d\alpha+\frac{1}{4\pi}ada-\frac{1}{2\pi}ad\alpha,
\end{equation}
where $\alpha$ is the gauge field that couples oppositely to $f_{1,2}$ and the four Dirac fermions occur at the Chern number changing transition. When $m<0$, integrating out the Dirac fermions gives the term $\frac{2}{4\pi}\alpha d\alpha$, which upon further integrating out $\alpha$, gives $\frac{3}{16\pi}ada$, consistent with $\nu=3/4$ FQH response for the holons that couple to $a$. When $m>0$,integrating out the Dirac fermions gives the term $-\frac{2}{4\pi}\alpha d\alpha$, which cancels the self Chern-Simons term for $\alpha$ and Higgs $a$. This indicates that bosons form a superfluid. The degeneracy of $4$ Dirac fermions at the transition is protected by the $1/4$ ``magnetic field" flux felt by $f_{i}$'s and essentially the filling. Whether this transition respects $C_6$ rotation remains to be investigated.

Combining the action for the spinon, we have for the complete critical theory for a transition from composite Fermi liquid at $\nu=3/4$ and Fermi liquid,
\begin{align}
    \mathcal L_{CFL-FL}=\mathcal L_{b;a,\alpha}+\mathcal L_{f_s;a}.\nonumber\\
\end{align}

Hence a generalization of the transition studied in this report to $\nu=3/4$ will be indicative of the putative composite Fermi liquid to Fermi liquid transition in moire TMD platforms.

%% file: arxiv.bbl
\begin{thebibliography}{38}%
\makeatletter
\providecommand \@ifxundefined [1]{%
 \@ifx{#1\undefined}
}%
\providecommand \@ifnum [1]{%
 \ifnum #1\expandafter \@firstoftwo
 \else \expandafter \@secondoftwo
 \fi
}%
\providecommand \@ifx [1]{%
 \ifx #1\expandafter \@firstoftwo
 \else \expandafter \@secondoftwo
 \fi
}%
\providecommand \natexlab [1]{#1}%
\providecommand \enquote  [1]{``#1''}%
\providecommand \bibnamefont  [1]{#1}%
\providecommand \bibfnamefont [1]{#1}%
\providecommand \citenamefont [1]{#1}%
\providecommand \href@noop [0]{\@secondoftwo}%
\providecommand \href [0]{\begingroup \@sanitize@url \@href}%
\providecommand \@href[1]{\@@startlink{#1}\@@href}%
\providecommand \@@href[1]{\endgroup#1\@@endlink}%
\providecommand \@sanitize@url [0]{\catcode `\\12\catcode `\$12\catcode
  `\&12\catcode `\#12\catcode `\^12\catcode `\_12\catcode `\%12\relax}%
\providecommand \@@startlink[1]{}%
\providecommand \@@endlink[0]{}%
\providecommand \url  [0]{\begingroup\@sanitize@url \@url }%
\providecommand \@url [1]{\endgroup\@href {#1}{\urlprefix }}%
\providecommand \urlprefix  [0]{URL }%
\providecommand \Eprint [0]{\href }%
\providecommand \doibase [0]{http://dx.doi.org/}%
\providecommand \selectlanguage [0]{\@gobble}%
\providecommand \bibinfo  [0]{\@secondoftwo}%
\providecommand \bibfield  [0]{\@secondoftwo}%
\providecommand \translation [1]{[#1]}%
\providecommand \BibitemOpen [0]{}%
\providecommand \bibitemStop [0]{}%
\providecommand \bibitemNoStop [0]{.\EOS\space}%
\providecommand \EOS [0]{\spacefactor3000\relax}%
\providecommand \BibitemShut  [1]{\csname bibitem#1\endcsname}%
\let\auto@bib@innerbib\@empty
\bibitem [{\citenamefont {Sachdev}(1999)}]{sachdev1999quantum}%
  \BibitemOpen
  \bibfield  {author} {\bibinfo {author} {\bibfnamefont {Subir}\ \bibnamefont
  {Sachdev}},\ }\bibfield  {title} {\enquote {\bibinfo {title} {Quantum phase
  transitions},}\ }\href@noop {} {\bibfield  {journal} {\bibinfo  {journal}
  {Physics world}\ }\textbf {\bibinfo {volume} {12}},\ \bibinfo {pages} {33}
  (\bibinfo {year} {1999})}\BibitemShut {NoStop}%
\bibitem [{\citenamefont {Senthil}(2024)}]{senthil2024deconfined}%
  \BibitemOpen
  \bibfield  {author} {\bibinfo {author} {\bibfnamefont {T}~\bibnamefont
  {Senthil}},\ }\bibfield  {title} {\enquote {\bibinfo {title} {Deconfined
  quantum critical points: a review},}\ }\href@noop {} {\bibfield  {journal}
  {\bibinfo  {journal} {50 Years of the Renormalization Group: Dedicated to the
  Memory of Michael E Fisher}\ ,\ \bibinfo {pages} {169--195}} (\bibinfo {year}
  {2024})}\BibitemShut {NoStop}%
\bibitem [{\citenamefont {Wen}(2004)}]{wen2004quantum}%
  \BibitemOpen
  \bibfield  {author} {\bibinfo {author} {\bibfnamefont {Xiao-Gang}\
  \bibnamefont {Wen}},\ }\href@noop {} {\emph {\bibinfo {title} {Quantum field
  theory of many-body systems: From the origin of sound to an origin of light
  and electrons}}}\ (\bibinfo  {publisher} {Oxford university press},\ \bibinfo
  {year} {2004})\BibitemShut {NoStop}%
\bibitem [{\citenamefont {Levin}\ and\ \citenamefont {Senthil}(2004)}]{DQCP1}%
  \BibitemOpen
  \bibfield  {author} {\bibinfo {author} {\bibfnamefont {Michael}\ \bibnamefont
  {Levin}}\ and\ \bibinfo {author} {\bibfnamefont {T.}~\bibnamefont
  {Senthil}},\ }\bibfield  {title} {\enquote {\bibinfo {title} {Deconfined
  quantum criticality and n\'eel order via dimer disorder},}\ }\href {\doibase
  10.1103/PhysRevB.70.220403} {\bibfield  {journal} {\bibinfo  {journal} {Phys.
  Rev. B}\ }\textbf {\bibinfo {volume} {70}},\ \bibinfo {pages} {220403}
  (\bibinfo {year} {2004})}\BibitemShut {NoStop}%
\bibitem [{\citenamefont {Senthil}\ \emph
  {et~al.}(2004{\natexlab{a}})\citenamefont {Senthil}, \citenamefont {Balents},
  \citenamefont {Sachdev}, \citenamefont {Vishwanath},\ and\ \citenamefont
  {Fisher}}]{DQCP2}%
  \BibitemOpen
  \bibfield  {author} {\bibinfo {author} {\bibfnamefont {T.}~\bibnamefont
  {Senthil}}, \bibinfo {author} {\bibfnamefont {Leon}\ \bibnamefont {Balents}},
  \bibinfo {author} {\bibfnamefont {Subir}\ \bibnamefont {Sachdev}}, \bibinfo
  {author} {\bibfnamefont {Ashvin}\ \bibnamefont {Vishwanath}}, \ and\ \bibinfo
  {author} {\bibfnamefont {Matthew P.~A.}\ \bibnamefont {Fisher}},\ }\bibfield
  {title} {\enquote {\bibinfo {title} {Quantum criticality beyond the
  landau-ginzburg-wilson paradigm},}\ }\href {\doibase
  10.1103/PhysRevB.70.144407} {\bibfield  {journal} {\bibinfo  {journal} {Phys.
  Rev. B}\ }\textbf {\bibinfo {volume} {70}},\ \bibinfo {pages} {144407}
  (\bibinfo {year} {2004}{\natexlab{a}})}\BibitemShut {NoStop}%
\bibitem [{\citenamefont {Senthil}\ \emph
  {et~al.}(2004{\natexlab{b}})\citenamefont {Senthil}, \citenamefont
  {Vishwanath}, \citenamefont {Balents}, \citenamefont {Sachdev},\ and\
  \citenamefont {Fisher}}]{DQCP3}%
  \BibitemOpen
  \bibfield  {author} {\bibinfo {author} {\bibfnamefont {T.}~\bibnamefont
  {Senthil}}, \bibinfo {author} {\bibfnamefont {Ashvin}\ \bibnamefont
  {Vishwanath}}, \bibinfo {author} {\bibfnamefont {Leon}\ \bibnamefont
  {Balents}}, \bibinfo {author} {\bibfnamefont {Subir}\ \bibnamefont
  {Sachdev}}, \ and\ \bibinfo {author} {\bibfnamefont {Matthew P.~A.}\
  \bibnamefont {Fisher}},\ }\bibfield  {title} {\enquote {\bibinfo {title}
  {Deconfined quantum critical points},}\ }\href {\doibase
  10.1126/science.1091806} {\bibfield  {journal} {\bibinfo  {journal}
  {Science}\ }\textbf {\bibinfo {volume} {303}},\ \bibinfo {pages} {1490--1494}
  (\bibinfo {year} {2004}{\natexlab{b}})},\ \Eprint
  {http://arxiv.org/abs/https://www.science.org/doi/pdf/10.1126/science.1091806}
  {https://www.science.org/doi/pdf/10.1126/science.1091806} \BibitemShut
  {NoStop}%
\bibitem [{\citenamefont {Burnell}(2018)}]{burnell2018anyon}%
  \BibitemOpen
  \bibfield  {author} {\bibinfo {author} {\bibfnamefont {Fiona~J}\ \bibnamefont
  {Burnell}},\ }\bibfield  {title} {\enquote {\bibinfo {title} {Anyon
  condensation and its applications},}\ }\href@noop {} {\bibfield  {journal}
  {\bibinfo  {journal} {Annual Review of Condensed Matter Physics}\ }\textbf
  {\bibinfo {volume} {9}},\ \bibinfo {pages} {307--327} (\bibinfo {year}
  {2018})}\BibitemShut {NoStop}%
\bibitem [{\citenamefont {Chubukov}\ \emph {et~al.}(1994)\citenamefont
  {Chubukov}, \citenamefont {Senthil},\ and\ \citenamefont
  {Sachdev}}]{XY_str1}%
  \BibitemOpen
  \bibfield  {author} {\bibinfo {author} {\bibfnamefont {Andrey~V.}\
  \bibnamefont {Chubukov}}, \bibinfo {author} {\bibfnamefont {T.}~\bibnamefont
  {Senthil}}, \ and\ \bibinfo {author} {\bibfnamefont {Subir}\ \bibnamefont
  {Sachdev}},\ }\bibfield  {title} {\enquote {\bibinfo {title} {Universal
  magnetic properties of frustrated quantum antiferromagnets in two
  dimensions},}\ }\href {\doibase 10.1103/PhysRevLett.72.2089} {\bibfield
  {journal} {\bibinfo  {journal} {Phys. Rev. Lett.}\ }\textbf {\bibinfo
  {volume} {72}},\ \bibinfo {pages} {2089--2092} (\bibinfo {year}
  {1994})}\BibitemShut {NoStop}%
\bibitem [{\citenamefont {Jalabert}\ and\ \citenamefont
  {Sachdev}(1991)}]{jalabert1991spontaneous}%
  \BibitemOpen
  \bibfield  {author} {\bibinfo {author} {\bibfnamefont {Rodolfo~A}\
  \bibnamefont {Jalabert}}\ and\ \bibinfo {author} {\bibfnamefont {Subir}\
  \bibnamefont {Sachdev}},\ }\bibfield  {title} {\enquote {\bibinfo {title}
  {Spontaneous alignment of frustrated bonds in an anisotropic,
  three-dimensional ising model},}\ }\href@noop {} {\bibfield  {journal}
  {\bibinfo  {journal} {Physical Review B}\ }\textbf {\bibinfo {volume} {44}},\
  \bibinfo {pages} {686} (\bibinfo {year} {1991})}\BibitemShut {NoStop}%
\bibitem [{\citenamefont {Senthil}\ and\ \citenamefont
  {Fisher}(2000)}]{senthil2000z}%
  \BibitemOpen
  \bibfield  {author} {\bibinfo {author} {\bibfnamefont {T}~\bibnamefont
  {Senthil}}\ and\ \bibinfo {author} {\bibfnamefont {Matthew~PA}\ \bibnamefont
  {Fisher}},\ }\bibfield  {title} {\enquote {\bibinfo {title} {Z 2 gauge theory
  of electron fractionalization in strongly correlated systems},}\ }\href@noop
  {} {\bibfield  {journal} {\bibinfo  {journal} {Physical Review B}\ }\textbf
  {\bibinfo {volume} {62}},\ \bibinfo {pages} {7850} (\bibinfo {year}
  {2000})}\BibitemShut {NoStop}%
\bibitem [{\citenamefont {Zhang}\ \emph {et~al.}(2023)\citenamefont {Zhang},
  \citenamefont {Zhu},\ and\ \citenamefont {Vishwanath}}]{XY*}%
  \BibitemOpen
  \bibfield  {author} {\bibinfo {author} {\bibfnamefont {Ya-Hui}\ \bibnamefont
  {Zhang}}, \bibinfo {author} {\bibfnamefont {Zheng}\ \bibnamefont {Zhu}}, \
  and\ \bibinfo {author} {\bibfnamefont {Ashvin}\ \bibnamefont {Vishwanath}},\
  }\bibfield  {title} {\enquote {\bibinfo {title} {Xy* transition and
  extraordinary boundary criticality from fractional exciton condensation in
  quantum hall bilayer},}\ }\href {\doibase 10.1103/PhysRevX.13.031023}
  {\bibfield  {journal} {\bibinfo  {journal} {Phys. Rev. X}\ }\textbf {\bibinfo
  {volume} {13}},\ \bibinfo {pages} {031023} (\bibinfo {year}
  {2023})}\BibitemShut {NoStop}%
\bibitem [{\citenamefont {Kivelson}\ \emph {et~al.}(1992)\citenamefont
  {Kivelson}, \citenamefont {Lee},\ and\ \citenamefont
  {Zhang}}]{kivelson1992global}%
  \BibitemOpen
  \bibfield  {author} {\bibinfo {author} {\bibfnamefont {Steven}\ \bibnamefont
  {Kivelson}}, \bibinfo {author} {\bibfnamefont {Dung-Hai}\ \bibnamefont
  {Lee}}, \ and\ \bibinfo {author} {\bibfnamefont {Shou-Cheng}\ \bibnamefont
  {Zhang}},\ }\bibfield  {title} {\enquote {\bibinfo {title} {Global phase
  diagram in the quantum hall effect},}\ }\href@noop {} {\bibfield  {journal}
  {\bibinfo  {journal} {Physical Review B}\ }\textbf {\bibinfo {volume} {46}},\
  \bibinfo {pages} {2223} (\bibinfo {year} {1992})}\BibitemShut {NoStop}%
\bibitem [{\citenamefont {Chen}\ \emph {et~al.}(1993)\citenamefont {Chen},
  \citenamefont {Fisher},\ and\ \citenamefont {Wu}}]{chen1993mott}%
  \BibitemOpen
  \bibfield  {author} {\bibinfo {author} {\bibfnamefont {Wei}\ \bibnamefont
  {Chen}}, \bibinfo {author} {\bibfnamefont {Matthew~PA}\ \bibnamefont
  {Fisher}}, \ and\ \bibinfo {author} {\bibfnamefont {Yong-Shi}\ \bibnamefont
  {Wu}},\ }\bibfield  {title} {\enquote {\bibinfo {title} {Mott transition in
  an anyon gas},}\ }\href@noop {} {\bibfield  {journal} {\bibinfo  {journal}
  {Physical Review B}\ }\textbf {\bibinfo {volume} {48}},\ \bibinfo {pages}
  {13749} (\bibinfo {year} {1993})}\BibitemShut {NoStop}%
\bibitem [{\citenamefont {Wen}\ and\ \citenamefont
  {Wu}(1993)}]{wen1993transitions}%
  \BibitemOpen
  \bibfield  {author} {\bibinfo {author} {\bibfnamefont {Xiao-Gang}\
  \bibnamefont {Wen}}\ and\ \bibinfo {author} {\bibfnamefont {Yong-Shi}\
  \bibnamefont {Wu}},\ }\bibfield  {title} {\enquote {\bibinfo {title}
  {Transitions between the quantum hall states and insulators induced by
  periodic potentials},}\ }\href@noop {} {\bibfield  {journal} {\bibinfo
  {journal} {Physical review letters}\ }\textbf {\bibinfo {volume} {70}},\
  \bibinfo {pages} {1501} (\bibinfo {year} {1993})}\BibitemShut {NoStop}%
\bibitem [{\citenamefont {Barkeshli}\ and\ \citenamefont
  {McGreevy}(2014)}]{MaissamLaughlin}%
  \BibitemOpen
  \bibfield  {author} {\bibinfo {author} {\bibfnamefont {Maissam}\ \bibnamefont
  {Barkeshli}}\ and\ \bibinfo {author} {\bibfnamefont {John}\ \bibnamefont
  {McGreevy}},\ }\bibfield  {title} {\enquote {\bibinfo {title} {Continuous
  transition between fractional quantum hall and superfluid states},}\ }\href
  {\doibase 10.1103/PhysRevB.89.235116} {\bibfield  {journal} {\bibinfo
  {journal} {Phys. Rev. B}\ }\textbf {\bibinfo {volume} {89}},\ \bibinfo
  {pages} {235116} (\bibinfo {year} {2014})}\BibitemShut {NoStop}%
\bibitem [{\citenamefont {Lee}\ \emph {et~al.}(2018)\citenamefont {Lee},
  \citenamefont {Wang}, \citenamefont {Zaletel}, \citenamefont {Vishwanath},\
  and\ \citenamefont {He}}]{JYL}%
  \BibitemOpen
  \bibfield  {author} {\bibinfo {author} {\bibfnamefont {Jong~Yeon}\
  \bibnamefont {Lee}}, \bibinfo {author} {\bibfnamefont {Chong}\ \bibnamefont
  {Wang}}, \bibinfo {author} {\bibfnamefont {Michael~P.}\ \bibnamefont
  {Zaletel}}, \bibinfo {author} {\bibfnamefont {Ashvin}\ \bibnamefont
  {Vishwanath}}, \ and\ \bibinfo {author} {\bibfnamefont {Yin-Chen}\
  \bibnamefont {He}},\ }\bibfield  {title} {\enquote {\bibinfo {title}
  {Emergent multi-flavor ${\mathrm{qed}}_{3}$ at the plateau transition between
  fractional chern insulators: Applications to graphene heterostructures},}\
  }\href {\doibase 10.1103/PhysRevX.8.031015} {\bibfield  {journal} {\bibinfo
  {journal} {Phys. Rev. X}\ }\textbf {\bibinfo {volume} {8}},\ \bibinfo {pages}
  {031015} (\bibinfo {year} {2018})}\BibitemShut {NoStop}%
\bibitem [{\citenamefont {Song}\ and\ \citenamefont
  {Zhang}(2023)}]{song_deconfined2023}%
  \BibitemOpen
  \bibfield  {author} {\bibinfo {author} {\bibfnamefont {Xue-Yang}\
  \bibnamefont {Song}}\ and\ \bibinfo {author} {\bibfnamefont {Ya-Hui}\
  \bibnamefont {Zhang}},\ }\bibfield  {title} {\enquote {\bibinfo {title}
  {{Deconfined criticalities and dualities between chiral spin liquid,
  topological superconductor and charge density wave Chern insulator}},}\
  }\href {\doibase 10.21468/SciPostPhys.15.5.215} {\bibfield  {journal}
  {\bibinfo  {journal} {SciPost Phys.}\ }\textbf {\bibinfo {volume} {15}},\
  \bibinfo {pages} {215} (\bibinfo {year} {2023})}\BibitemShut {NoStop}%
\bibitem [{\citenamefont {Song}\ \emph {et~al.}(2024)\citenamefont {Song},
  \citenamefont {Zhang},\ and\ \citenamefont {Senthil}}]{song_2024_transition}%
  \BibitemOpen
  \bibfield  {author} {\bibinfo {author} {\bibfnamefont {Xue-Yang}\
  \bibnamefont {Song}}, \bibinfo {author} {\bibfnamefont {Ya-Hui}\ \bibnamefont
  {Zhang}}, \ and\ \bibinfo {author} {\bibfnamefont {T.}~\bibnamefont
  {Senthil}},\ }\bibfield  {title} {\enquote {\bibinfo {title} {Phase
  transitions out of quantum hall states in moir\'e materials},}\ }\href
  {\doibase 10.1103/PhysRevB.109.085143} {\bibfield  {journal} {\bibinfo
  {journal} {Phys. Rev. B}\ }\textbf {\bibinfo {volume} {109}},\ \bibinfo
  {pages} {085143} (\bibinfo {year} {2024})}\BibitemShut {NoStop}%
\bibitem [{\citenamefont {Barkeshli}\ and\ \citenamefont
  {McGreevy}(2012)}]{MaissamCFL}%
  \BibitemOpen
  \bibfield  {author} {\bibinfo {author} {\bibfnamefont {Maissam}\ \bibnamefont
  {Barkeshli}}\ and\ \bibinfo {author} {\bibfnamefont {John}\ \bibnamefont
  {McGreevy}},\ }\bibfield  {title} {\enquote {\bibinfo {title} {Continuous
  transitions between composite fermi liquid and landau fermi liquid: A route
  to fractionalized mott insulators},}\ }\href {\doibase
  10.1103/PhysRevB.86.075136} {\bibfield  {journal} {\bibinfo  {journal} {Phys.
  Rev. B}\ }\textbf {\bibinfo {volume} {86}},\ \bibinfo {pages} {075136}
  (\bibinfo {year} {2012})}\BibitemShut {NoStop}%
\bibitem [{Note1()}]{Note1}%
  \BibitemOpen
  \bibinfo {note} {A $1/N_f$ expansion predicts conformality, but its validity
  at $N_f = 2$ is uncertain.}\BibitemShut {Stop}%
\bibitem [{\citenamefont {Barkeshli}\ \emph {et~al.}(2015)\citenamefont
  {Barkeshli}, \citenamefont {Yao},\ and\ \citenamefont
  {Laumann}}]{NormLaughlin}%
  \BibitemOpen
  \bibfield  {author} {\bibinfo {author} {\bibfnamefont {M.}~\bibnamefont
  {Barkeshli}}, \bibinfo {author} {\bibfnamefont {N.~Y.}\ \bibnamefont {Yao}},
  \ and\ \bibinfo {author} {\bibfnamefont {C.~R.}\ \bibnamefont {Laumann}},\
  }\bibfield  {title} {\enquote {\bibinfo {title} {Continuous preparation of a
  fractional chern insulator},}\ }\href {\doibase
  10.1103/PhysRevLett.115.026802} {\bibfield  {journal} {\bibinfo  {journal}
  {Phys. Rev. Lett.}\ }\textbf {\bibinfo {volume} {115}},\ \bibinfo {pages}
  {026802} (\bibinfo {year} {2015})}\BibitemShut {NoStop}%
\bibitem [{\citenamefont {Kamal}\ \emph {et~al.}(2022)\citenamefont {Kamal},
  \citenamefont {Kemp}, \citenamefont {He}, \citenamefont {Fuji}, \citenamefont
  {Aidelsburger}, \citenamefont {Zoller},\ and\ \citenamefont
  {Yao}}]{NormChern}%
  \BibitemOpen
  \bibfield  {author} {\bibinfo {author} {\bibfnamefont {Helia}\ \bibnamefont
  {Kamal}}, \bibinfo {author} {\bibfnamefont {Jack}\ \bibnamefont {Kemp}},
  \bibinfo {author} {\bibfnamefont {Yin-Chen}\ \bibnamefont {He}}, \bibinfo
  {author} {\bibfnamefont {Yohei}\ \bibnamefont {Fuji}}, \bibinfo {author}
  {\bibfnamefont {Monika}\ \bibnamefont {Aidelsburger}}, \bibinfo {author}
  {\bibfnamefont {Peter}\ \bibnamefont {Zoller}}, \ and\ \bibinfo {author}
  {\bibfnamefont {Norman}\ \bibnamefont {Yao}},\ }\bibfield  {title} {\enquote
  {\bibinfo {title} {Floquet flux attachment in cold atomic systems},}\
  }\href@noop {} {\  (\bibinfo {year} {2022})},\ \Eprint
  {http://arxiv.org/abs/2401.08754} {arXiv:2401.08754 [quant-ph]} \BibitemShut
  {NoStop}%
\bibitem [{\citenamefont {Motruk}\ and\ \citenamefont
  {Pollmann}(2017)}]{Frank2017}%
  \BibitemOpen
  \bibfield  {author} {\bibinfo {author} {\bibfnamefont {Johannes}\
  \bibnamefont {Motruk}}\ and\ \bibinfo {author} {\bibfnamefont {Frank}\
  \bibnamefont {Pollmann}},\ }\bibfield  {title} {\enquote {\bibinfo {title}
  {Phase transitions and adiabatic preparation of a fractional chern insulator
  in a boson cold-atom model},}\ }\href {\doibase 10.1103/PhysRevB.96.165107}
  {\bibfield  {journal} {\bibinfo  {journal} {Phys. Rev. B}\ }\textbf {\bibinfo
  {volume} {96}},\ \bibinfo {pages} {165107} (\bibinfo {year}
  {2017})}\BibitemShut {NoStop}%
\bibitem [{\citenamefont {Zeng}(2021)}]{Zeng2021}%
  \BibitemOpen
  \bibfield  {author} {\bibinfo {author} {\bibfnamefont {Tian-Sheng}\
  \bibnamefont {Zeng}},\ }\bibfield  {title} {\enquote {\bibinfo {title} {Phase
  transitions of bosonic fractional quantum hall effect in topological flat
  bands},}\ }\href {\doibase 10.1103/PhysRevB.103.085122} {\bibfield  {journal}
  {\bibinfo  {journal} {Phys. Rev. B}\ }\textbf {\bibinfo {volume} {103}},\
  \bibinfo {pages} {085122} (\bibinfo {year} {2021})}\BibitemShut {NoStop}%
\bibitem [{SM()}]{SM}%
  \BibitemOpen
  \href@noop {} {}\bibinfo {note} {For more details see the Supplementary
  Information.}\BibitemShut {Stop}%
\bibitem [{\citenamefont {Haldane}(1983)}]{ParentHamiltonian}%
  \BibitemOpen
  \bibfield  {author} {\bibinfo {author} {\bibfnamefont {F.~D.~M.}\
  \bibnamefont {Haldane}},\ }\bibfield  {title} {\enquote {\bibinfo {title}
  {Fractional quantization of the hall effect: A hierarchy of incompressible
  quantum fluid states},}\ }\href {\doibase 10.1103/PhysRevLett.51.605}
  {\bibfield  {journal} {\bibinfo  {journal} {Phys. Rev. Lett.}\ }\textbf
  {\bibinfo {volume} {51}},\ \bibinfo {pages} {605--608} (\bibinfo {year}
  {1983})}\BibitemShut {NoStop}%
\bibitem [{\citenamefont {Zaletel}\ and\ \citenamefont
  {Mong}(2012)}]{ExactMPS}%
  \BibitemOpen
  \bibfield  {author} {\bibinfo {author} {\bibfnamefont {Michael~P.}\
  \bibnamefont {Zaletel}}\ and\ \bibinfo {author} {\bibfnamefont {Roger S.~K.}\
  \bibnamefont {Mong}},\ }\bibfield  {title} {\enquote {\bibinfo {title} {Exact
  matrix product states for quantum hall wave functions},}\ }\href {\doibase
  10.1103/PhysRevB.86.245305} {\bibfield  {journal} {\bibinfo  {journal} {Phys.
  Rev. B}\ }\textbf {\bibinfo {volume} {86}},\ \bibinfo {pages} {245305}
  (\bibinfo {year} {2012})}\BibitemShut {NoStop}%
\bibitem [{\citenamefont {Zaletel}\ \emph {et~al.}(2013)\citenamefont
  {Zaletel}, \citenamefont {Mong},\ and\ \citenamefont {Pollmann}}]{TopoChara}%
  \BibitemOpen
  \bibfield  {author} {\bibinfo {author} {\bibfnamefont {Michael~P.}\
  \bibnamefont {Zaletel}}, \bibinfo {author} {\bibfnamefont {Roger S.~K.}\
  \bibnamefont {Mong}}, \ and\ \bibinfo {author} {\bibfnamefont {Frank}\
  \bibnamefont {Pollmann}},\ }\bibfield  {title} {\enquote {\bibinfo {title}
  {Topological characterization of fractional quantum hall ground states from
  microscopic hamiltonians},}\ }\href {\doibase 10.1103/PhysRevLett.110.236801}
  {\bibfield  {journal} {\bibinfo  {journal} {Phys. Rev. Lett.}\ }\textbf
  {\bibinfo {volume} {110}},\ \bibinfo {pages} {236801} (\bibinfo {year}
  {2013})}\BibitemShut {NoStop}%
\bibitem [{\citenamefont {Seidel}\ and\ \citenamefont
  {Lee}(2006)}]{lee_abelian}%
  \BibitemOpen
  \bibfield  {author} {\bibinfo {author} {\bibfnamefont {Alexander}\
  \bibnamefont {Seidel}}\ and\ \bibinfo {author} {\bibfnamefont {Dung-Hai}\
  \bibnamefont {Lee}},\ }\bibfield  {title} {\enquote {\bibinfo {title}
  {Abelian and non-abelian hall liquids and charge-density wave: Quantum number
  fractionalization in one and two dimensions},}\ }\href {\doibase
  10.1103/PhysRevLett.97.056804} {\bibfield  {journal} {\bibinfo  {journal}
  {Phys. Rev. Lett.}\ }\textbf {\bibinfo {volume} {97}},\ \bibinfo {pages}
  {056804} (\bibinfo {year} {2006})}\BibitemShut {NoStop}%
\bibitem [{Note2()}]{Note2}%
  \BibitemOpen
  \bibinfo {note} {For very narrow cylinders the Laughlin state evolves into a
  charge-density wave~\cite {lee_abelian}. At the larger \(L_y\) used here the
  entanglement spectrum confirms genuine Laughlin order.}\BibitemShut {Stop}%
\bibitem [{\citenamefont {Li}\ and\ \citenamefont
  {Haldane}(2008)}]{Li-Haldane}%
  \BibitemOpen
  \bibfield  {author} {\bibinfo {author} {\bibfnamefont {Hui}\ \bibnamefont
  {Li}}\ and\ \bibinfo {author} {\bibfnamefont {F.~D.~M.}\ \bibnamefont
  {Haldane}},\ }\bibfield  {title} {\enquote {\bibinfo {title} {Entanglement
  spectrum as a generalization of entanglement entropy: Identification of
  topological order in non-abelian fractional quantum hall effect states},}\
  }\href {\doibase 10.1103/PhysRevLett.101.010504} {\bibfield  {journal}
  {\bibinfo  {journal} {Phys. Rev. Lett.}\ }\textbf {\bibinfo {volume} {101}},\
  \bibinfo {pages} {010504} (\bibinfo {year} {2008})}\BibitemShut {NoStop}%
\bibitem [{\citenamefont {Zeng}\ \emph {et~al.}(2020)\citenamefont {Zeng},
  \citenamefont {Sheng},\ and\ \citenamefont {Zhu}}]{zeng_2020_continuous}%
  \BibitemOpen
  \bibfield  {author} {\bibinfo {author} {\bibfnamefont {Tian-Sheng}\
  \bibnamefont {Zeng}}, \bibinfo {author} {\bibfnamefont {D.~N.}\ \bibnamefont
  {Sheng}}, \ and\ \bibinfo {author} {\bibfnamefont {W.}~\bibnamefont {Zhu}},\
  }\bibfield  {title} {\enquote {\bibinfo {title} {Continuous phase transition
  between bosonic integer quantum hall liquid and a trivial insulator: Evidence
  for deconfined quantum criticality},}\ }\href {\doibase
  10.1103/PhysRevB.101.035138} {\bibfield  {journal} {\bibinfo  {journal}
  {Phys. Rev. B}\ }\textbf {\bibinfo {volume} {101}},\ \bibinfo {pages}
  {035138} (\bibinfo {year} {2020})}\BibitemShut {NoStop}%
\bibitem [{\citenamefont {He}\ \emph {et~al.}(2017)\citenamefont {He},
  \citenamefont {Zaletel}, \citenamefont {Oshikawa},\ and\ \citenamefont
  {Pollmann}}]{MikeDSL}%
  \BibitemOpen
  \bibfield  {author} {\bibinfo {author} {\bibfnamefont {Yin-Chen}\
  \bibnamefont {He}}, \bibinfo {author} {\bibfnamefont {Michael~P.}\
  \bibnamefont {Zaletel}}, \bibinfo {author} {\bibfnamefont {Masaki}\
  \bibnamefont {Oshikawa}}, \ and\ \bibinfo {author} {\bibfnamefont {Frank}\
  \bibnamefont {Pollmann}},\ }\bibfield  {title} {\enquote {\bibinfo {title}
  {Signatures of dirac cones in a dmrg study of the kagome heisenberg model},}\
  }\href {\doibase 10.1103/PhysRevX.7.031020} {\bibfield  {journal} {\bibinfo
  {journal} {Phys. Rev. X}\ }\textbf {\bibinfo {volume} {7}},\ \bibinfo {pages}
  {031020} (\bibinfo {year} {2017})}\BibitemShut {NoStop}%
\bibitem [{\citenamefont {Zauner}\ \emph {et~al.}(2015)\citenamefont {Zauner},
  \citenamefont {Draxler}, \citenamefont {Vanderstraeten}, \citenamefont
  {Degroote}, \citenamefont {Haegeman}, \citenamefont {Rams}, \citenamefont
  {Stojevic}, \citenamefont {Schuch},\ and\ \citenamefont
  {Verstraete}}]{transfer_matrix}%
  \BibitemOpen
  \bibfield  {author} {\bibinfo {author} {\bibfnamefont {V}~\bibnamefont
  {Zauner}}, \bibinfo {author} {\bibfnamefont {D}~\bibnamefont {Draxler}},
  \bibinfo {author} {\bibfnamefont {L}~\bibnamefont {Vanderstraeten}}, \bibinfo
  {author} {\bibfnamefont {M}~\bibnamefont {Degroote}}, \bibinfo {author}
  {\bibfnamefont {J}~\bibnamefont {Haegeman}}, \bibinfo {author} {\bibfnamefont
  {M~M}\ \bibnamefont {Rams}}, \bibinfo {author} {\bibfnamefont
  {V}~\bibnamefont {Stojevic}}, \bibinfo {author} {\bibfnamefont
  {N}~\bibnamefont {Schuch}}, \ and\ \bibinfo {author} {\bibfnamefont
  {F}~\bibnamefont {Verstraete}},\ }\bibfield  {title} {\enquote {\bibinfo
  {title} {Transfer matrices and excitations with matrix product states},}\
  }\href {\doibase 10.1088/1367-2630/17/5/053002} {\bibfield  {journal}
  {\bibinfo  {journal} {New Journal of Physics}\ }\textbf {\bibinfo {volume}
  {17}},\ \bibinfo {pages} {053002} (\bibinfo {year} {2015})}\BibitemShut
  {NoStop}%
\bibitem [{\citenamefont {Lu}\ \emph {et~al.}(2024{\natexlab{a}})\citenamefont
  {Lu}, \citenamefont {Wu}, \citenamefont {Chen},\ and\ \citenamefont
  {Meng}}]{ZiYangLaughlin}%
  \BibitemOpen
  \bibfield  {author} {\bibinfo {author} {\bibfnamefont {Hongyu}\ \bibnamefont
  {Lu}}, \bibinfo {author} {\bibfnamefont {Han-Qing}\ \bibnamefont {Wu}},
  \bibinfo {author} {\bibfnamefont {Bin-Bin}\ \bibnamefont {Chen}}, \ and\
  \bibinfo {author} {\bibfnamefont {Zi~Yang}\ \bibnamefont {Meng}},\ }\bibfield
   {title} {\enquote {\bibinfo {title} {Continuous transition between bosonic
  fractional chern insulator and superfluid},}\ }\href@noop {} {\  (\bibinfo
  {year} {2024}{\natexlab{a}})},\ \Eprint {http://arxiv.org/abs/2405.18269}
  {arXiv:2405.18269 [cond-mat.str-el]} \BibitemShut {NoStop}%
\bibitem [{\citenamefont {Lu}\ \emph {et~al.}(2024{\natexlab{b}})\citenamefont
  {Lu}, \citenamefont {Wu}, \citenamefont {Chen},\ and\ \citenamefont
  {Meng}}]{ZiYangLaughlin2}%
  \BibitemOpen
  \bibfield  {author} {\bibinfo {author} {\bibfnamefont {Hongyu}\ \bibnamefont
  {Lu}}, \bibinfo {author} {\bibfnamefont {Han-Qing}\ \bibnamefont {Wu}},
  \bibinfo {author} {\bibfnamefont {Bin-Bin}\ \bibnamefont {Chen}}, \ and\
  \bibinfo {author} {\bibfnamefont {Zi~Yang}\ \bibnamefont {Meng}},\ }\bibfield
   {title} {\enquote {\bibinfo {title} {Vestigial gapless boson density wave
  emerging between $\nu = 1/2$ fractional chern insulator and finite-momentum
  supersolid},}\ }\href@noop {} {\  (\bibinfo {year} {2024}{\natexlab{b}})},\
  \Eprint {http://arxiv.org/abs/2408.07111} {arXiv:2408.07111
  [cond-mat.mes-hall]} \BibitemShut {NoStop}%
\bibitem [{\citenamefont {Lotri{\v c}}\ and\ \citenamefont
  {Simon}(2025)}]{Lotric2025-mt}%
  \BibitemOpen
  \bibfield  {author} {\bibinfo {author} {\bibfnamefont {Tev{\v z}}\
  \bibnamefont {Lotri{\v c}}}\ and\ \bibinfo {author} {\bibfnamefont
  {Steven~H}\ \bibnamefont {Simon}},\ }\bibfield  {title} {\enquote {\bibinfo
  {title} {Paired parton trial states for the superfluid-fractional chern
  insulator transition},}\ }\href@noop {} {\  (\bibinfo {year} {2025})},\
  \Eprint {http://arxiv.org/abs/2504.20139} {arXiv:2504.20139
  [cond-mat.str-el]} \BibitemShut {NoStop}%
\bibitem [{\citenamefont {Dong}\ \emph {et~al.}(2023)\citenamefont {Dong},
  \citenamefont {Wang}, \citenamefont {Ledwith}, \citenamefont {Vishwanath},\
  and\ \citenamefont {Parker}}]{dong2023_composite}%
  \BibitemOpen
  \bibfield  {author} {\bibinfo {author} {\bibfnamefont {Junkai}\ \bibnamefont
  {Dong}}, \bibinfo {author} {\bibfnamefont {Jie}\ \bibnamefont {Wang}},
  \bibinfo {author} {\bibfnamefont {Patrick~J.}\ \bibnamefont {Ledwith}},
  \bibinfo {author} {\bibfnamefont {Ashvin}\ \bibnamefont {Vishwanath}}, \ and\
  \bibinfo {author} {\bibfnamefont {Daniel~E.}\ \bibnamefont {Parker}},\
  }\bibfield  {title} {\enquote {\bibinfo {title} {Composite fermi liquid at
  zero magnetic field in twisted ${\mathrm{mote}}_{2}$},}\ }\href {\doibase
  10.1103/PhysRevLett.131.136502} {\bibfield  {journal} {\bibinfo  {journal}
  {Phys. Rev. Lett.}\ }\textbf {\bibinfo {volume} {131}},\ \bibinfo {pages}
  {136502} (\bibinfo {year} {2023})}\BibitemShut {NoStop}%
\end{thebibliography}
